\documentclass[10pt,preprint]{aastex6}
\usepackage{natbib}
\usepackage{multirow}
\usepackage{upgreek}
\usepackage{wasysym}
\usepackage{txfonts}
\usepackage{gensymb}
\usepackage{morefloats}
\usepackage{threeparttable}
\slugcomment{version \today}
\shorttitle{X-ray emission from the nuclear region of Arp 220}
\shortauthors{Paggi et al.}
\begin{document}
\title{X-ray emission from the nuclear region of Arp 220}
\author{Alessandro Paggi\altaffilmark{1}, Giuseppina Fabbiano\altaffilmark{1}, Guido Risaliti\altaffilmark{1,2}, Junfeng Wang\altaffilmark{3}, Margarita Karovska\altaffilmark{1}, Martin Elvis\altaffilmark{1}, W. Peter Maksym\altaffilmark{1}, Jonathan McDowell\altaffilmark{1} and Jay Gallagher\altaffilmark{4}}
\affil{\altaffilmark{1}Harvard-Smithsonian Center for Astrophysics, 60 Garden St, Cambridge, MA 02138, USA: \href{mailto:apaggi@cfa.harvard.edu}{apaggi@cfa.harvard.edu}\\
\altaffilmark{2}INAF-Arcetri Observatory, Largo E, Fermi 5, I-50125 Firenze, Italy\\
\altaffilmark{3}Department of Astronomy and Institute of Theoretical Physics and Astrophysics, Xiamen University, Xiamen, China 361005\\
\altaffilmark{4}Department of Astronomy, University of Wisconsin, Madison, WI 53706-1582, USA}
\begin{abstract}
We present an imaging and spectral analysis of the nuclear region of the ULIRG merger Arp 220, using deep \textit{Chandra}-ACIS observations summing up to \(\sim 300\mbox{ ks}\). Narrow-band imaging with sub-pixel resolution of the innermost nuclear region reveals two distinct Fe-K emitting sources, coincident with the infrared and radio nuclear clusters. These sources are separated by 1' (\(\sim 380\) pc). {The X-ray emission is extended and elongated in the eastern nucleus, like the disk emission observed in millimeter radio images, suggesting starburst dominance in this region.} We estimate Fe-K equivalent width \(\gtrsim 1\) keV for both sources, and observed 2-10 keV luminosities \(\sim 2\times{10}^{40}\mbox{ erg}\mbox{ s}^{-1}\) (W) and \(\sim 3 \times {10}^{40}\mbox{ erg}\mbox{ s}^{-1}\) (E). In the 6-7 keV band the emission from these regions is dominated by the 6.7 keV Fe \textsc{xxv} line, suggesting contribution from collisionally ionized gas. The thermal energy content of this gas is consistent with kinetic energy injection in the interstellar medium by Type II SNe. However, nuclear winds from hidden AGN (\(\varv\sim 2000 \mbox{ km}\mbox{ s}^{-1}\)) cannot be excluded. The \(3\sigma\) upper limits on the neutral Fe-K\(\alpha\) flux of the nuclear regions correspond to intrinsic AGN 2-10 keV luminosities \(< 1\times {10}^{42}\mbox{ erg}\mbox{ s}^{-1}\) (W) and \(< 0.4\times {10}^{42}\mbox{ erg}\mbox{ s}^{-1}\) (E). For typical AGN SEDs the bolometric luminosities are \(< 3\times {10}^{43}\mbox{ erg}\mbox{ s}^{-1}\) (W) and \(< 8\times {10}^{43}\mbox{ erg}\mbox{ s}^{-1}\) (E), and black hole masses \(<1\times{10}^5 M_{\astrosun}\) (W) and \(< 5\times{10}^5 M_{\astrosun}\) (E) for Eddington limited AGNs with a standard 10\% efficiency.
\end{abstract}


\section{INTRODUCTION}\label{intro}
At a distance of 80 Mpc \citep{1998ApJS..119...41K}, Arp 220 (IC 4553/4) is both a merger, and the nearest Ultra-luminous IR Galaxy \citep[ULIRG;][]{1987ApJ...320..238S, 1996ARA&A..34..749S}. Near IR high-resolution (0.1'') NICMOS-HST imaging identifies the two nuclear regions of the merging galaxies, which are coincident with the two components of a double radio source \citep{1995ApJ...454..745B, 1998ApJ...492L.107S}. At a separation of 1'' (380 pc at a distance of 80 Mpc \footnote{In the following, we adopt the standard flat cosmology with \(\Omega_\Lambda = 0.69\) and \(H_0 = 68 \mbox{ km}\mbox{ s}^{-1}\mbox{ Mpc}^{-1}\) \citep{2016A&A...594A..13P}.}), these nuclei are closer together than the nuclei of NGC 6240 (\(\sim 690\) pc separation, \citealt{2003ApJ...582L..15K}), and therefore may be subject to even stronger gravitational interaction, leading to accretion on the nuclei of the merging galaxy \citep[e.g.,][]{1994MNRAS.271..317G, 2007Sci...316.1874M}.

The presence of AGN emission in addition to intense star formation in the two nuclei has been debated. It is also not clear where this AGN emission may reside. AGN contribution to the bolometric luminosity between \(\sim 5\%\) and \(\sim 25\%\) (\(0.1-0.4\times{10}^{12} L_{\astrosun}\)) is suggested by the Spitzer Mid-IR spectrum of the central 8" region of Arp 220 (\citealt{2009ApJS..182..628V, 2010MNRAS.405.2505N}; more recent \textit{Herschel} results \citep{2011ApJ...743...94R} and modeling of the nuclear spectra \citep{2013MNRAS.429..242C} agree with this conclusion. In the west nucleus of Arp 220 the presence of a maser \citep{2009A&A...493..481A} and a rotating massive molecular disk \citep{2007A&A...468L..57D} suggests a massive nuclear black hole. Instead, IRAM-PdBI observations \citep{2012ApJ...754...58K} showed in the eastern nucleus of Arp 220 an extended structure in the mm-CO gas emission, elongated in the NE-SW direction, and suggested a highly dust-obscured nucleus lying between the two NIR sources detected by \citet{1998ApJ...492L.107S}. Recently, an analysis of 3.5 and 1.2 mm IRAM-PdBI data performed by \citet{2015ApJ...800...25T} provided evidence of significant chemical differences between the two nuclei. These authors argue for a significant AGN contribution to the W nucleus luminosity, and starburst dominance in the E nucleus. ALMA imaging at 350 and 696 GHz \citep{2015ApJ...800...70S} suggested compact nuclear disks with masses of \(\sim 4\) and \(\sim 2\times {10}^9\,M_{\astrosun}\) for W and E nucleus, respectively, within a radius \(\sim 70\mbox{ pc}\). High-resolution VLA observations at 33 and 6 GHz by \citet{2015ApJ...799...10B} showed double radio morphology. These authors estimated for both nuclei high hydrogen column densities \(\sim {10}^{25}\mbox{ cm}^{-2}\), and the  star formation rate surface densities among the most extreme measured for any star-forming system (\(\sim {10}^4\,M_{\astrosun}\mbox{ yr}^{-1}\mbox{ kpc}^{-2}\)), without any compelling evidence of an AGN dominating the nuclei emission at 33 GHz.

The M-\(\sigma\) relation \citep[e.g.,][]{1998AJ....115.2285M} has suggested that the evolution of galaxies and super-massive nuclear black holes are linked. Both the stellar population and the SMBH of a galaxy are thought to grow and evolve by merging of smaller gas-rich galaxies and their nuclear SMBHs \citep{2005Natur.433..604D, 2008ApJ...682L..13H}. During this process, the SMBH may be ``buried'' by thick molecular gas, which feeds the SMBH at high rates, causing the birth of an obscured Compton Thick  \citep[CT,][]{1999ApJ...522..157R, 2006ApJ...648..111L} Active Galactic Nucleus (AGN). In this paper we re-examine the X-ray emission of Arp 220. A high resolution study of the archival \(\sim 57\) ks \textit{Chandra} ACIS observation \citep[obsid 869;][]{2002ApJ...581..974C} measured the hard X-ray luminosity of the central AGN but had insufficient signal to noise ratio and spatial resolution to detect a line or to spatially untangle the complex central emission. The X-ray spectrum extracted from the central \(\sim 3\)'' region suggested hard continuum and \(\sim 6.6\) keV Fe-K emission line. A subsequent \textit{XMM-Newton} observation detected Fe-K line emission centered at 6.7 keV with an equivalent width EW \(\sim 1.9\) keV \citep{2005MNRAS.357..565I} suggesting highly photoionized, low-density gas illuminated by a hidden CT AGN. A reanalysis of the \textit{Chandra} and \textit{XMM-Newton} data \citep{2011ApJ...729...52L} only managed to set an upper limit on the neutral Fe-K\(\alpha\) emission at 6.4 keV.

Here we make use of new deep \textit{Chandra}-ACIS data. Together with the archival observation, we reach a total exposure of \(\sim 300\mbox{ ks}\), which allows us to perform a study of the X-ray emission from the nuclear region of Arp 220 with unprecedented detail, by means of sub-pixel imaging of \textit{Chandra} ACIS data in narrow spectral ranges. This technique has been used successfully to study crowded emission regions of nearby Seyferts (e.g. in NGC 4151 \citealt{2011ApJ...729...75W, 2011ApJ...736...62W, 2011ApJ...742...23W}; Mrk 573, \citealt{2012ApJ...756...39P}). Our new look at the nuclear region of Arp 220, has resulted in the discovery of two sources in the 6-7 keV Fe-K band, spatially coincident with the near-IR and radio positions. In Section \ref{data} we present the data reduction procedures for the imaging and spectral analysis; in Section \ref{discussion} we discuss and interpret our results; in Section \ref{conclusions} we then draw our conclusions. In the following analysis we fix the Galactic absorption to the value \(N_{H, gal}= {3.9}\times{10}^{20}\mbox{ cm}^{-2}\).

\section{Data Reduction}\label{data}

\textit{Chandra} observations used in this analysis are listed in Table \ref{tab:observations}. Level 2 event files were retrieved from the \textit{Chandra} Data Archive\footnote{\href{http://cda.harvard.edu/chaser}{http://cda.harvard.edu/chaser}} and reduced with the CIAO (\citealt{2006SPIE.6270E..60F}) 4.7 software and the \textit{Chandra} Calibration Data Base (\textsc{caldb}) 4.6.7, adopting standard procedures. After excluding time intervals of background flares exceeding \(3\sigma\) with the \textsc{lc\_sigma\_clip} task, we obtained a low-background total exposure time of \(\sim 290\) ks. The nucleus has no significant pile up, as measured by the CIAO \textsc{pileup\_map} tool\footnote{{Pile-up occurs on X-ray CCDs for source with high flux levels, when two or more photons arrive within the same detector region within a single CCD frame integration time, and they are counted a single photon of higher energy\citep{2001ApJ...562..575D}.}}.

\subsection{Imaging analysis}\label{imaging_analysis}

Imaging analysis was performed without pixel randomization to take advantage of the telescope dithering in event positioning and with the sub-pixel event repositioning (SER) procedure \citep{2003ApJ...590..586L}. We used a pixel size 1/4 of \(0.492"\), the native \textit{Chandra}/ACIS detector pixel \citep[see, e.g.,][]{2004ApJ...615..161H, 2007ApJ...657..145S, 2010ApJ...708..171P, 2011ApJ...729...75W}. Using the same Orion ACIS-S data as in the calibration of \citet{2003ApJ...590..586L}, we find a significant \(\Delta=50\%\) (improvement in PSF FWHM as defined in \citealt{2003ApJ...590..586L}) from sub-pixel repositioning for an on-axis source at 6-7 keV (\(\Delta=70\%\) at \(\sim 2\) keV because of the narrower PSF at larger energies). Most of the imaging improvement is from sub-pixel event repositioning (SER) procedure \citep{2003ApJ...590..586L} and without pixel randomization, to take advantage of the sampling of the PSF by the well characterized spacecraft dither motion. Because of the similarly `peaked' inner PSF, this is similarly effective at 2 and 6 keV.

\begin{figure}
\begin{center}
\includegraphics[scale=0.95]{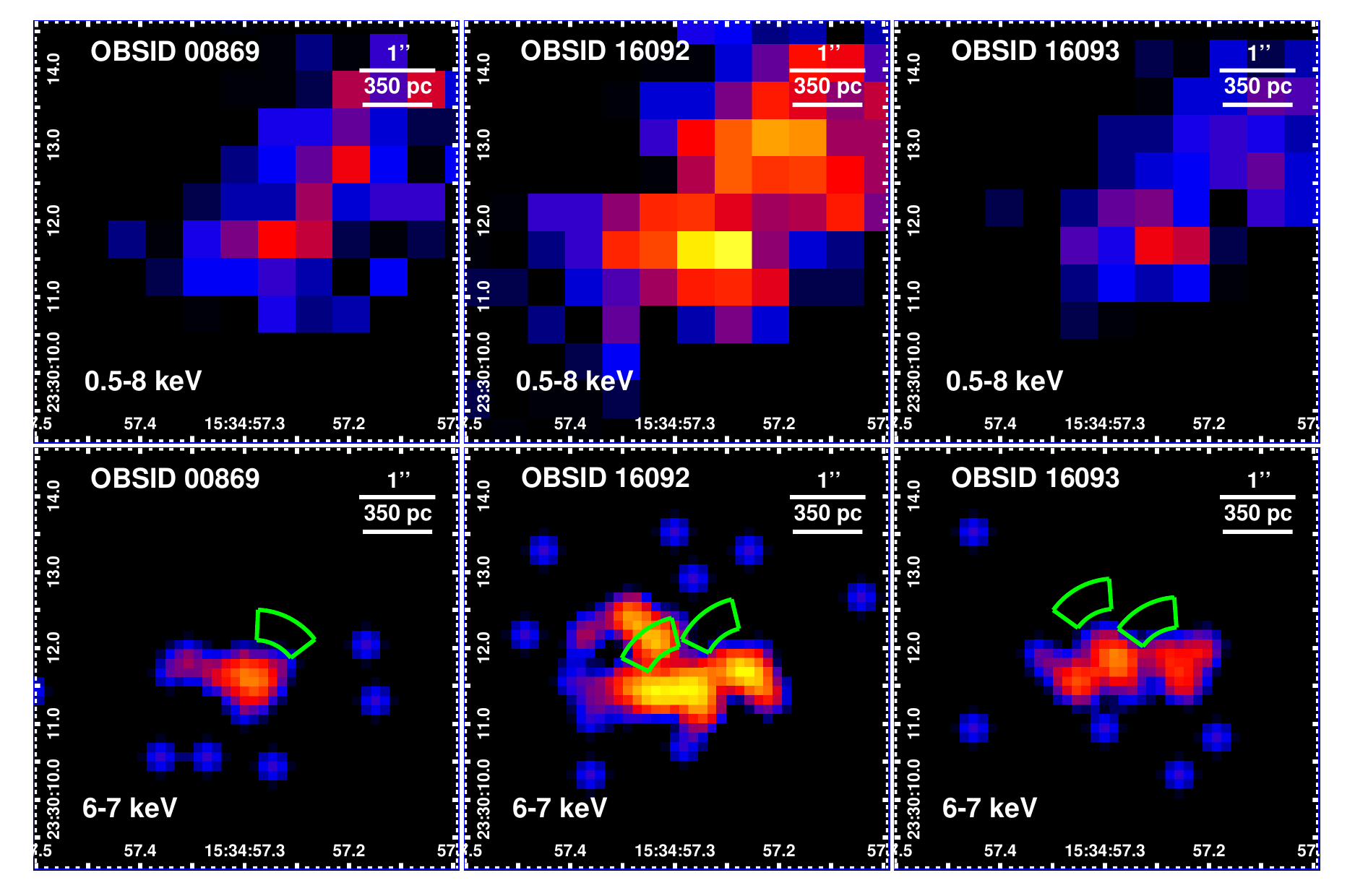}
\caption{(Top panels) Broad band 0.5-8 keV images of the central \(\sim 5''\) region of Arp 220 for three \textit{Chandra}-ACIS observations reprojected on OBSID 16092, with native \(0.492''\) pixel size. (Bottom panels) Same as top panels but in narrow band 6-7 keV. Images are presented with sub-pixel binning (1/4 of the native pixel size) and 3X3 pixel FWHM gaussian filter smoothing. Green regions show the location of the \textit{Chandra} PSF artifacts for the two nuclei.}\label{fig:reprojection}
\end{center}
\end{figure}

In the top panels of Figure \ref{fig:reprojection} we show the central 5'' region of Arp 220 as imaged by the three \textit{Chandra}-ACIS observations in the broad 0.5-8 keV band with native \(0.492''\) pixel size. To merge the three exposures we first used the \textsc{wavdetect} task to identify point sources in the field of each observation (excluding the central region). We then used the \textsc{reproject\_aspect} task to modify the aspect solutions minimizing position differences between these sources and finally merged the re-aligned images with the \textsc{reproject\_events} task to reproject the event files to the deepest 16092 observation. The relative shifts between the observations is \(\sim 0.5''\), comparable with the \(0.6''\) \textit{Chandra} astrometric uncertainty. 

In the bottom panels of Figure \ref{fig:reprojection} the same central 5'' region is shown in the narrow Fe-K 6-7 keV band with sub-pixel binning of 1/4 of the native pixel size and 3 pixel FWHM gaussian filter smoothing. The narrow-band images for the deeper observations - 16092 and 16093 - show two sources associated with the West (W) and East (E) Arp 220 nuclei separated by 1'' (\(\sim 380\) pc at the source distance), while in the shorter 00869 observation only the W nucleus is detected. The locations of these unique emission regions strongly argue for an identification of these sources with the nuclei of the merging galaxies. On the same panels we also show in green the position of the \textit{Chandra} PSF artifacts as obtained with \textsc{make\_psf\_asymmetry\_region}\footnote{\href{http://cxc.harvard.edu/ciao/ahelp/make\_psf\_asymmetry\_region.html}{http://cxc.harvard.edu/ciao/ahelp/make\_psf\_asymmetry\_region.html}} tool for the two nuclei. PSF asymmetries are expected in the north-west direction for Obs ID 869, and in north-east direction for Obs IDs 16092 and 16093 (due to different roll angles). The two nuclear sources are not affected by PSF asymmetries, but the extension in the north-east direction of the E source, shown in the deepest observation 16092, overlaps with the region of \textit{Chandra} PSF asymmetry which, however, can account for only up to 10\% of the counts of this feature. The resulting merged images are shown in Figure \ref{fig:counts} in the 3-6 (left panel), {and 6.4-6.7 (central panel)} keV bands with sub-pixel binning 1/4 of the native pixel size. {In addition, in the right panel of the same figure we show the 6.4-6.7 keV band image after continuum (3-6 keV) subtraction (see Sect. \ref{spectral_analysis}).} These images are indicative {of continuum and Fe \textsc{xxv} line emission}. The regions of W and E nuclei considered for spectral extraction discussed in Section \ref{spectral_analysis} are shown as white circles.

Using PSF simulations performed with \textit{Chandra} Ray Tracer (ChaRT\footnote{\href{http://cxc.harvard.edu/chart/}{http://cxc.harvard.edu/chart/}}, \citealt{2003ASPC..295..477C}) taking into account the source spectrum, exposure time and off-axis angle, we applied the Expectation through Markov Chain Monte Carlo \citep{2004ApJ...610.1213E, 2005ApJ...623L.137K, 2007ApJ...661.1048K, 2014ApJ...781...55W} PSF-deconvolution algorithm to the merged images of the narrow-band emission. This method yields a multi-scale image reconstruction specifically applicable to Poisson noise limited data.

\begin{figure}
\centering
\includegraphics[scale=0.9]{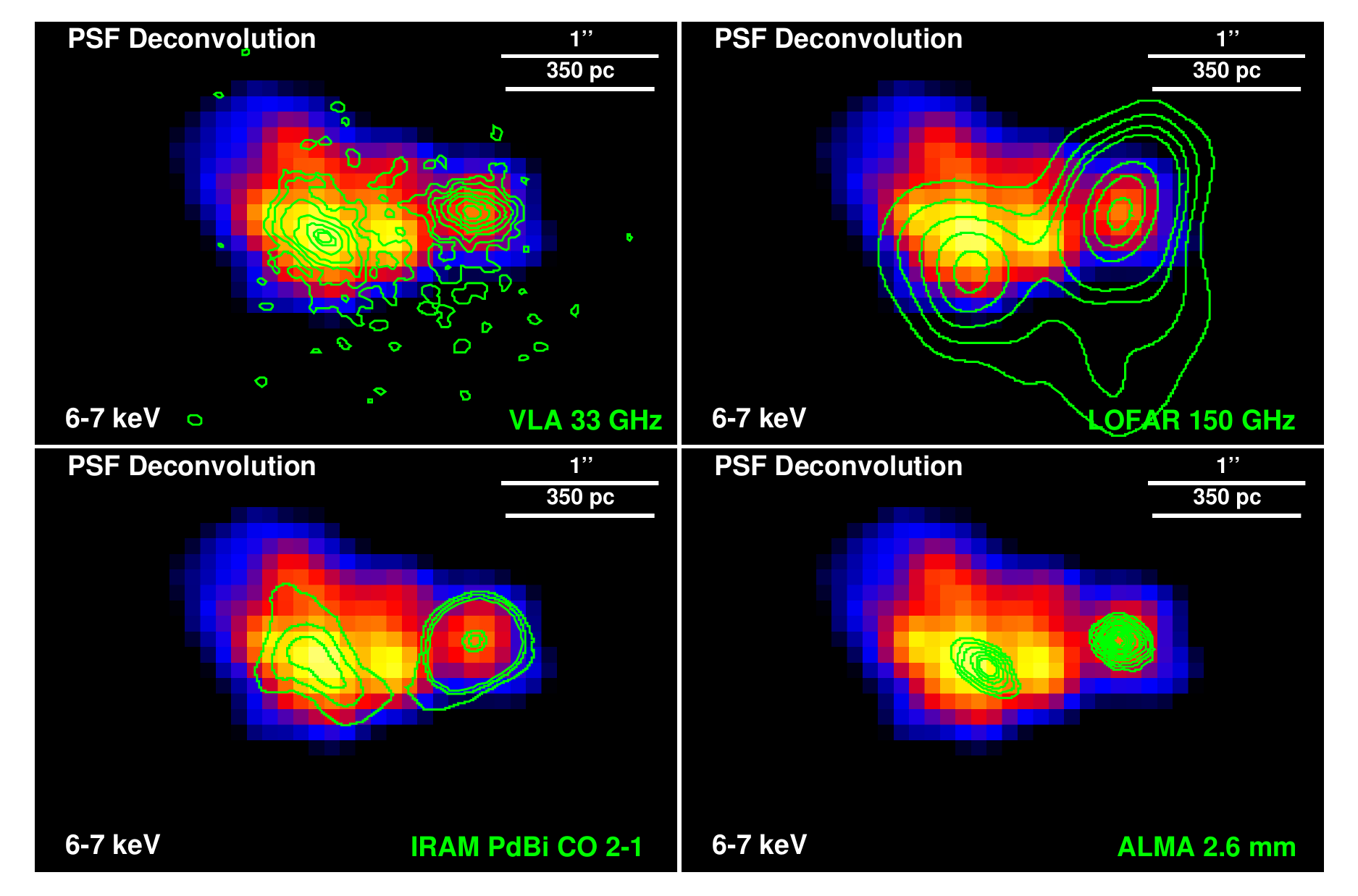}
\caption{Merged, PSF-deconvolved \textit{Chandra}-ACIS 6-7 keV image with sub-pixel binning 1/5 of the native pixel size and 3 pixel FWHM gaussian filter smoothing, with superimposed (from top left to bottom right) contours from 33 GHz VLA data \citep{2015ApJ...799...10B}, the 150 MHz continuum and CO 2-1 {IRAM PdBi} data \citep{2016A&A...593A..86V}, and the 2.6 mm continuum ALMA data \citep{2017ApJ...836...66S}, registered as discussed in the main text.}\label{fig:deconvolution}
\end{figure}

As noted by \citet{2002ApJ...581..974C}, the lack of USNO detected stars in Arp 220 field does not allow  accurate absolute astrometry using field stars. However, a comparison between the merged 6-7 keV \textit{Chandra}-ACIS image and the high-resolution 33 GHz VLA observations \citep{2015ApJ...799...10B} shows a clear similarity in the morphology of the nuclei. 
{In order to highlight the morphological similarity between the X-ray and the radio emission, we shifted the} VLA 33 GHz sources shifting {the latter} in the NE direction by \(\sim 0.2"\) (compatible with the \textit{Chandra} astrometric accuracy, \citealt{2011ApJS..192....8R}) in order to match the position of the western radio lobe with the western X-ray nucleus, {that appears less extended and allows therefore a more accurate positioning}. The result of this registration procedure is shown in Figure \ref{fig:deconvolution}. In both the X-ray image and the radio contours the W nucleus looks compact while the E nucleus appears somewhat extended. We performed a similar registration with radio contours of 150 MHz {LOFAR} continuum and CO 2-1 {IRAM PdBi} images \citep{2016A&A...593A..86V}, and with 2.6 mm continuum ALMA images \citep{2017ApJ...836...66S}. {Again we matched the western nuclei. We notice that the positions of western lobes in VLA 33 GHz, LOFAR 150 MHz, IRAM PdBi CO 2-1 and ALMA 2.6 mm images differ less \(\sim 0.1''\), and therefore the use of these maps yields similar results. In all cases the} deconvolved X-ray image shows a more compact emission from the W nucleus, and a more extended E nucleus with a morphology similar to that of the 33 GHz, CO 2-1 and 2.6 mm radio emissions {suggesting starburst dominance in this region. On the other hand,} the E X-ray nucleus lies somewhat north of the E lobe observed in the 150 MHz continuum LOFAR image. The central peak corresponds with the X4 X-ray source from \citet{2002ApJ...581..974C}. 
{Since the registration of radio images to the X-ray data yields similar results, in the following we will compare to VLA 33 GHz data.}

\begin{figure}
\centering
\includegraphics[scale=0.9]{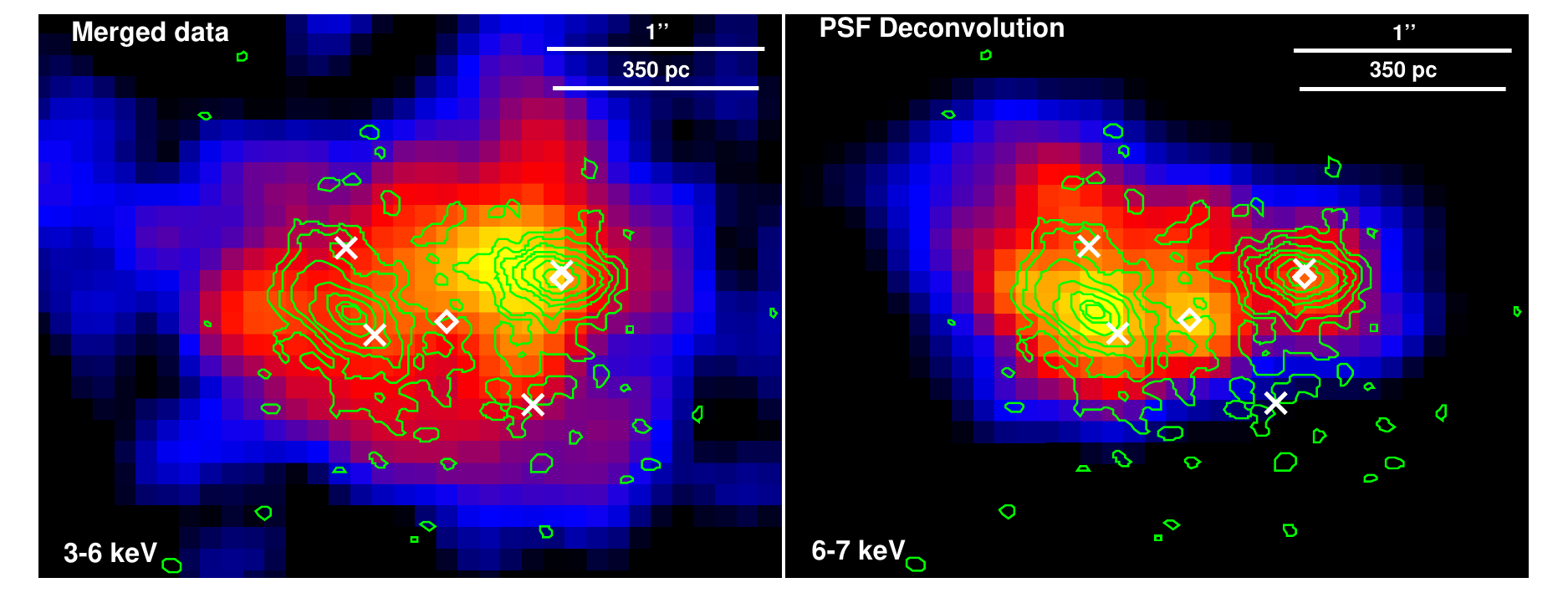}
\caption{Merged \textit{Chandra}-ACIS {3-6 keV (left panel) and 6-7 keV PSF-deconvolved (right panel)} images, indicative of continuum and Fe line emission, respectively. Images are presented with sub-pixel binning 1/5 of the native pixel size and 3 pixel FWHM gaussian filter smoothing. Green lines represent 33 GHz VLA contours as in Figure \ref{fig:deconvolution}, white crosses represent the NIR sources  from \citet{1998ApJ...492L.107S} and  white diamonds represent the X-ray sources from \citet{2002ApJ...581..974C}.}\label{fig:sources}
\end{figure}

Figure \ref{fig:sources} compares the \textit{Chandra} merged images in the \(3-6\) keV continuum with the observed 6-7 keV band selected to represent both the 6.4 keV emitted Fe K-\(\alpha\) neutral line and the 6.7 keV Fe \textsc{xxv} line. The 3-6 keV continuum emission centroid lies in between the peaks of the nuclear line emission (although closer to the W nucleus). The 6-7 keV peak coincide with the NIR sources reported by \citet{2002ApJ...581..974C}, and the 5 GHz peaks. The Fe 6-7 keV image suggests a more extended E nucleus emission with respect to the W one. The X1 source reported by \citet{2002ApJ...581..974C} in the first 57 ks \textit{Chandra} dataset coincides with the W nucleus, while the X4 source from the same study lies between the two nuclei and it is coincident with the central peak seen the PSF-deconvolved 6-7 keV image.

\subsection{Spectral analysis}\label{spectral_analysis}

We attempted a spectral characterization of the emission, extracting 3-8 keV spectra from the two circular regions indicated in Figure \ref{fig:counts}. The two easternmost peaks shown in Figure \ref{fig:deconvolution} do not have enough counts to be studied separately (see Table \ref{tab:fits}) and we therefore selected a region that encompasses them both.
Spectra were extracted with \textsc{CIAO specextract} task, applying a point-source aperture correction, binned to obtain a minimum of \(5\) counts per bin, and fitted employing the Cash statistic.

\subsection{AGN Model}

\begin{figure}
\centering
\includegraphics[scale=0.9]{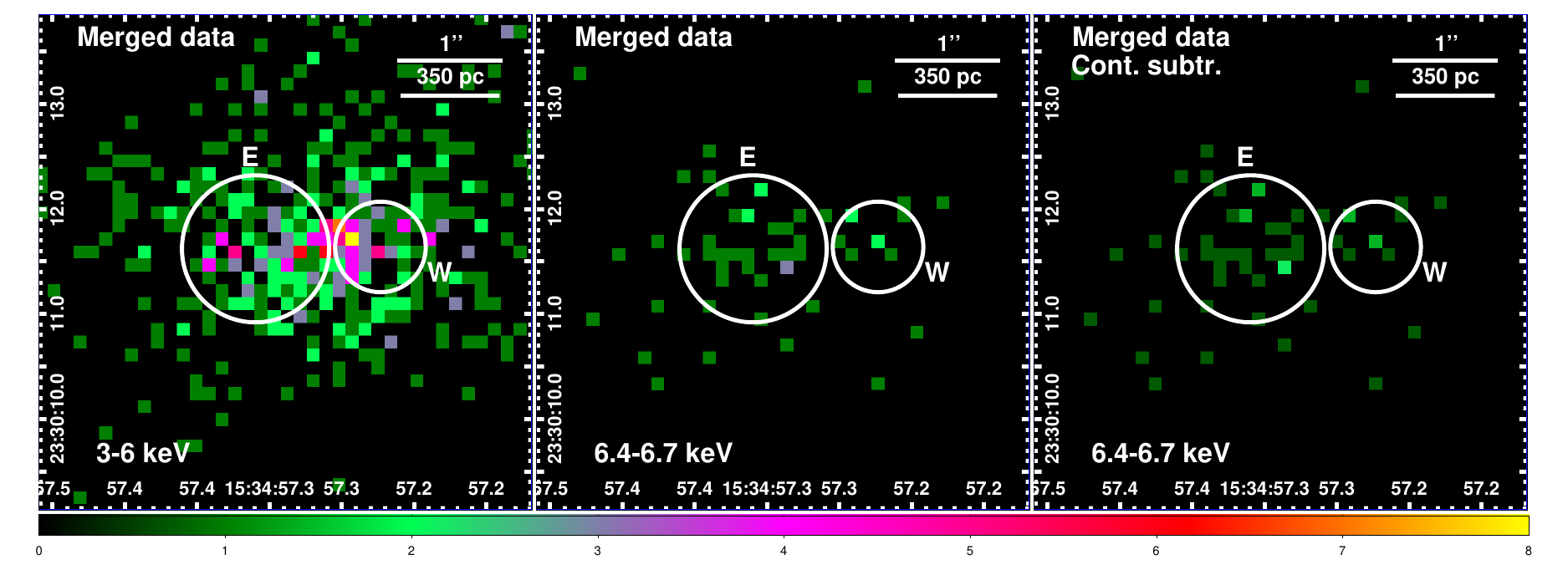}
\caption{From left to right: merged \textit{Chandra}-ACIS {3-6 and 6.4-6.7} keV images, indicative of {continuum and Fe \textsc{xxv}} line emission, respectively. {In the right panel we show the 6.4-6.7 keV band image after continuum (3-6 keV) subtraction.} Images are presented with sub-pixel binning 1/4 of the native pixel size. The regions of spectral extraction considered in Section \ref{spectral_analysis} are shown as white circles.}\label{fig:counts}
\end{figure}

\begin{figure}
\centering
\includegraphics[angle=-90,scale=0.29]{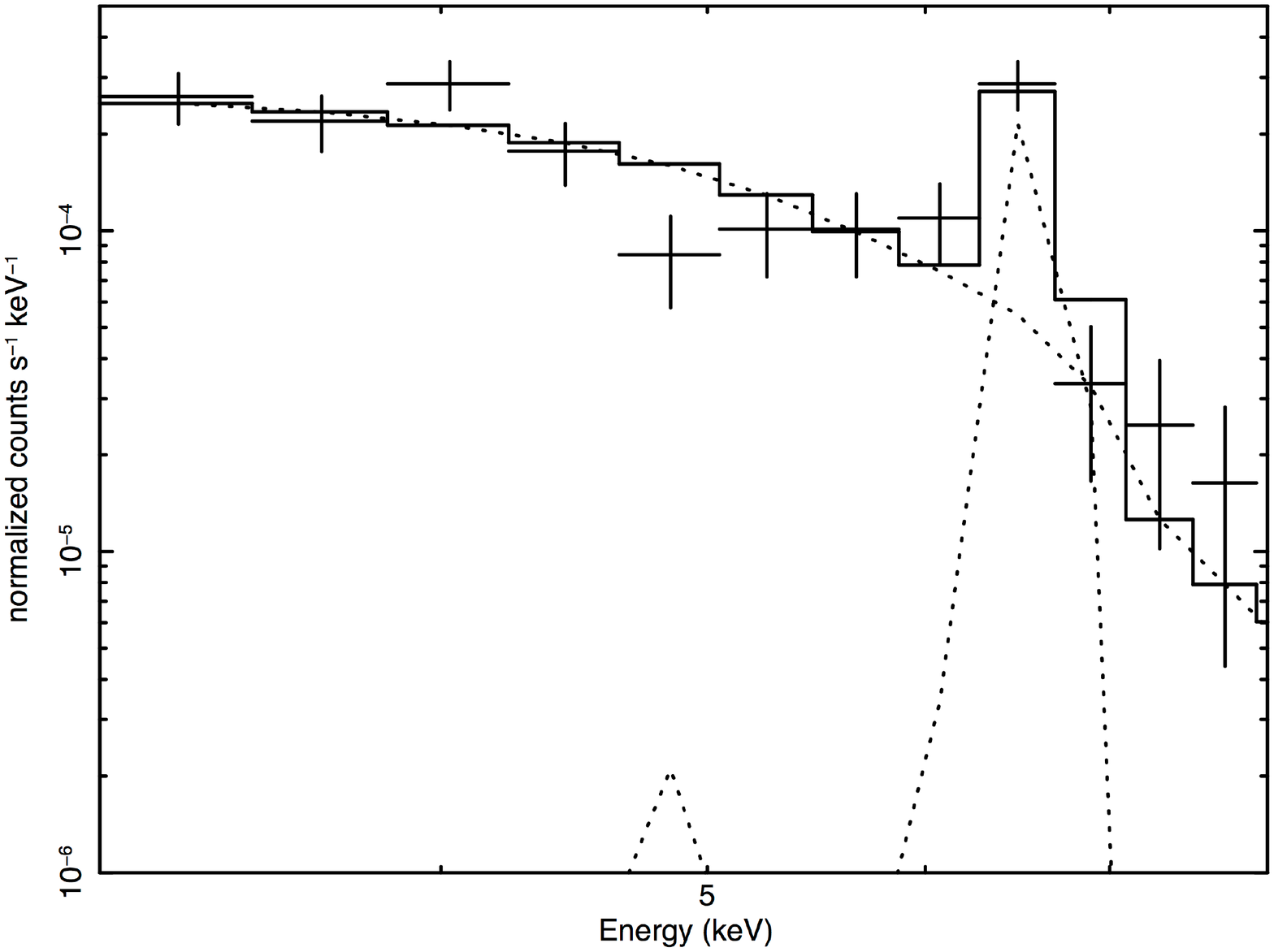}
\includegraphics[angle=-90,scale=0.29]{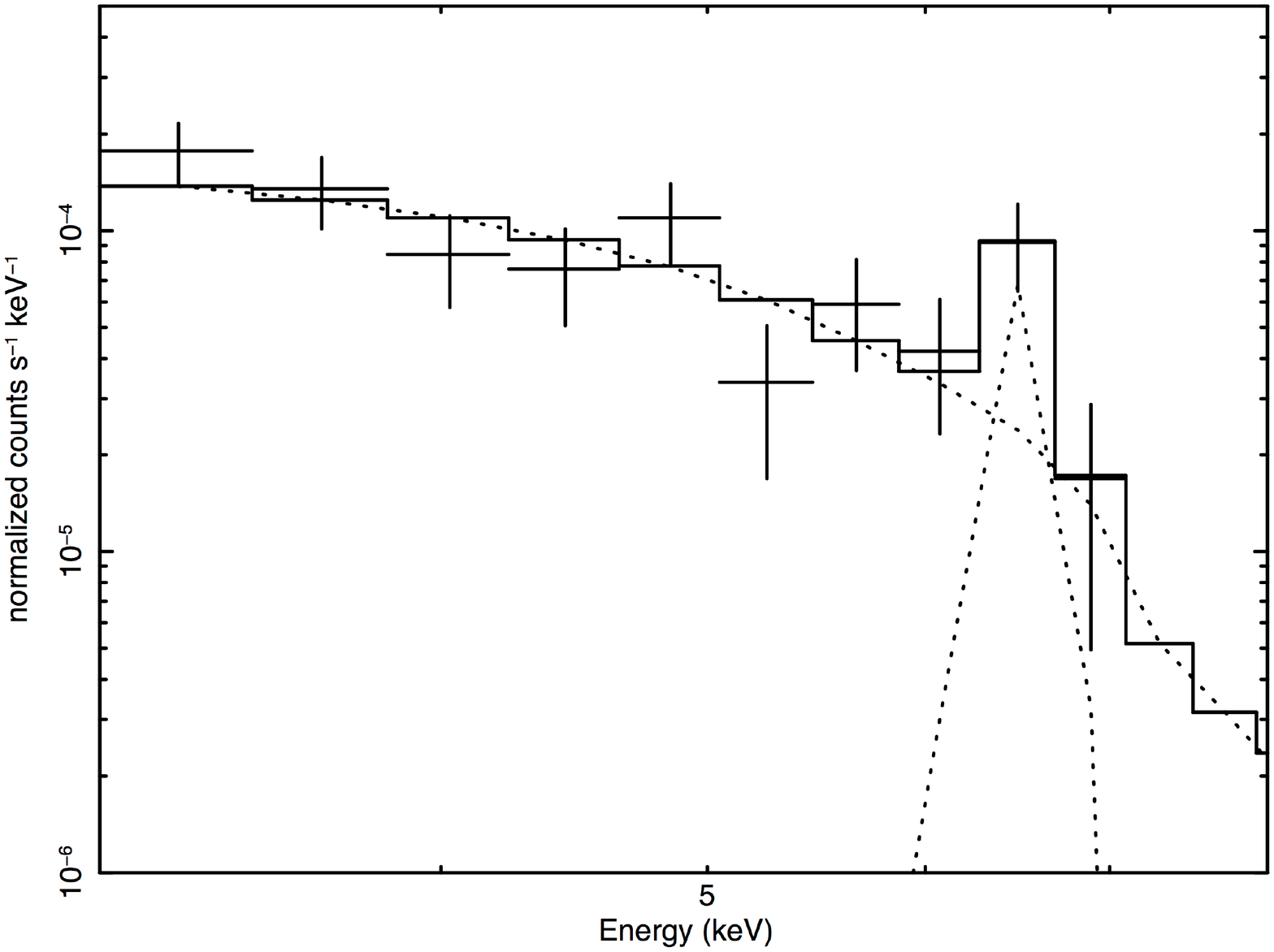}
\caption{Best fit to spectra extracted from regions shown in Figure \ref{fig:counts} from E region (left panel) and W region (right panel). For clarity only the reflection component and Fe-K line are shown with a dotted line.}\label{fig:spectra}
\end{figure}

We first used a model typical of CT AGN emission \citep{2006ApJ...648..111L}, comprising an absorption component fixed to the Galactic value \({3.9}\times{10}^{20}\mbox{ cm}^{-2}\), a power-law and red-shifted gaussian Fe-K line(s) with width fixed to 100 eV, plus possible additional Ar, S and Ca lines as reported by \citet{2005MNRAS.357..565I}. We used both \textsc{XSPEC} (ver. 12.8.2\footnote{\href{https://heasarc.gsfc.nasa.gov/xanadu/xspec}{https://heasarc.gsfc.nasa.gov/xanadu/xspec}}) and \textsc{Sherpa}\footnote{\href{http://cxc.harvard.edu/sherpa} {http://cxc.harvard.edu/sherpa}} with identical results. The extracted spectra and the best-fit parameters are presented in Figure \ref{fig:spectra} and Table \ref{tab:fits}\footnote{In the following, errors correspond to the \(1\)-\(\sigma\) confidence level for one parameter of interest.}, respectively. Spectra were extracted from nuclear regions shown in Figure \ref{fig:counts}, as well as in the entire central region of Arp 220 using a circular region of 4.5" radius centered at the coordinates of sources shown in Fig \ref{fig:counts}.

Due to low statistics, we first fit the power-law component excluding data from 6-7 keV energy range. We then froze the power-law spectral index so obtained and added a red-shifted gaussian Fe-K line with energy free to vary (columns 1, 4 and 7 of Table \ref{tab:fits}, model a). In order to evaluate the contribution from neutral and ionized Fe-K separately, we then froze the line rest-frame energy at 6.4 keV (columns 2, 5 and 8 of Table \ref{tab:fits}, model b), and then added a second line with rest-frame energy frozen at 6.7 keV (columns 3, 6 and 9 of Table \ref{tab:fits}, model c) in order to evaluate the relative contribution of neutral and ionized Fe lines.

In each model, given the contamination of the continuum radiation by the extended emission (suggested by the spatial distributions of continuum photons in Figure \ref{fig:counts}) and the poor statistics, our estimate of the nuclear continuum luminosity is an upper limit, and the Fe-K EWs must be considered as lower limits.

\begin{figure}
\centering
\includegraphics[scale=0.4]{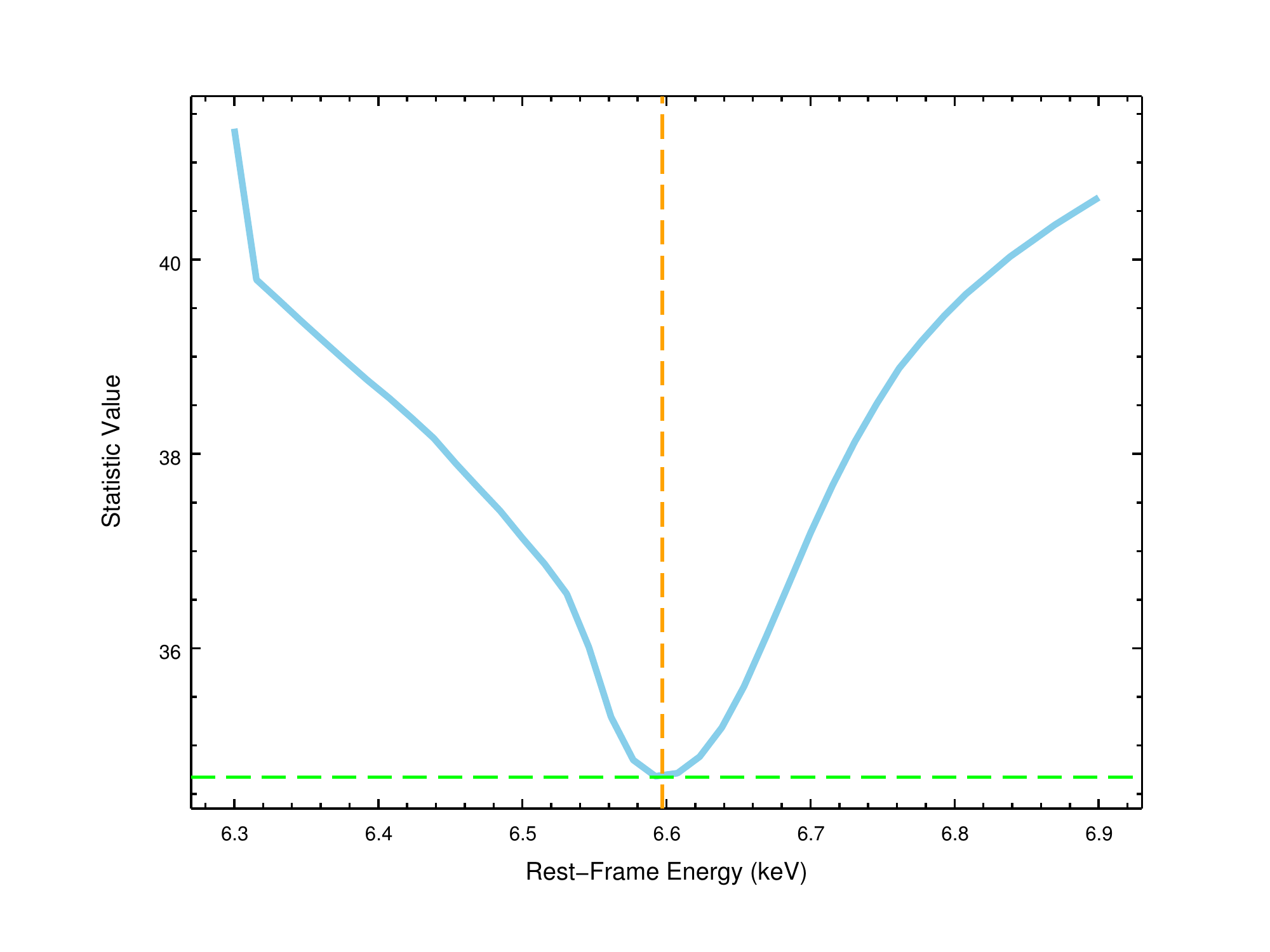}
\includegraphics[scale=0.4]{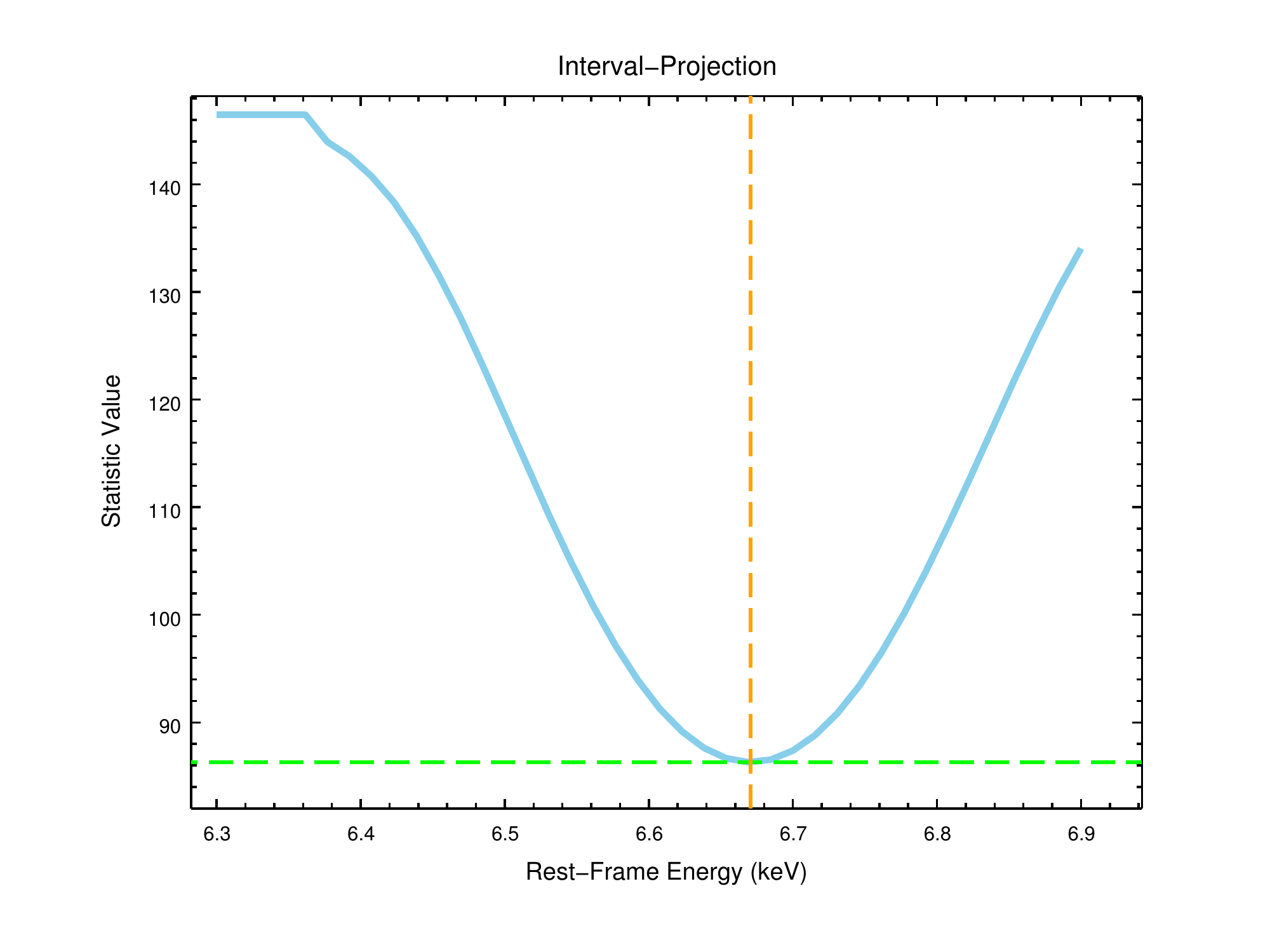}
\caption{Value of the fit statistic of model a as function of the rest-frame line energy in W (left panel) and E (right panel) nuclei.}\label{fig:chi2}
\end{figure}

\begin{figure}
\centering
\includegraphics[scale=0.3]{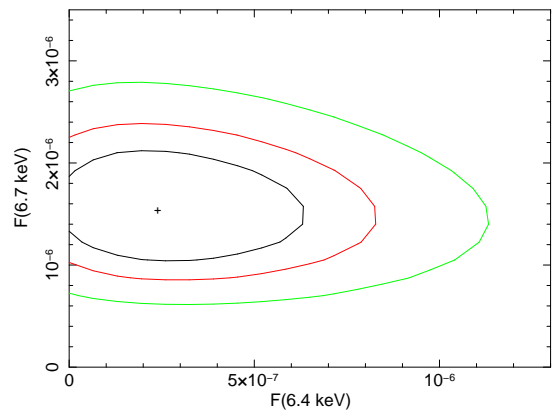}
\caption{Contour plot for the relative flux of the 6.4 and 6.7 keV lines in a fit where both rest-frame energies are fixed in the \textit{XMM-Newton} data form the nuclear (E+W) region. The contours represent a probability of 68\%, 90\% and 99\%. Even if a solution with a single 6.7 keV line is the more likely, a contribution of up to 40\% by a neutral line is compatible with the data at a 90\% confidence level.}\label{fig:contours}
\end{figure}

\begin{itemize}
\item[Model (a)] We added a single gaussian line to the power law. This leads to detections of  Fe-K line features in both regions, with comparable EW (1.2 keV and 1.8 keV in W and E region, respectively). As expected from the imaging, the Fe-K line is more luminous in the E nucleus (\(0.8\times{10}^{40}\mbox{ erg}\mbox{ s}^{-1}\)) with respect to the {W} (\(0.3\times{10}^{40}\mbox{ erg}\mbox{ s}^{-1}\)). As already discussed, the \textit{observed} 2-10 keV luminosities, \(3.2\times{10}^{40}\mbox{ erg}\mbox{ s}^{-1}\) ({E}) and \(2.0\times{10}^{40}\mbox{ erg}\mbox{ s}^{-1}\) ({W}), should be considered as upper limits because of contamination. Given the rest-frame line energies, ({\({6.67}\pm{0.03}\) and \({6.61}_{-0.07}^{+0.06}\)}keV in {E} and {W} regions, respectively) Fe \textsc{xxv} appears to be dominant in the Fe-K range, with fluxes of \({9.43}_{-1.91}^{+2.17} \times{10}^{-7}\mbox{ cm}^{-2}\mbox{ s}^{-1}\) and \({4.01}_{-1.59}^{+1.67}\) {in E and W regions}, respectively. In Figure \ref{fig:chi2} we show the value of the fit statistic as a function of the rest-frame line energy.

We compare our results with those of \citet{2011ApJ...729...52L}, who extracted a spectrum of the entire central region of Arp 220 using the archival \textit{Chandra}-ACIS observation (OBSID 00869). Using a circular 4.5" radius count extraction region we detect a line at \({6.65}\pm{0.02}\) keV with and equivalent width of \({1.39}_{-0.69}^{+1.44}\) keV, compatible with the \textit{XMM-Newton} detection at \({6.72}_{-0.12}^{+0.10}\) keV \citep{2005MNRAS.357..565I}.
\item[Model b)] To assess an upper limit to the neutral Fe-K\(\alpha\) contribution, we held the line rest-frame energy fixed at 6.4 keV (model b). Using a circular 4.5" radius extraction region like \citet{2011ApJ...729...52L}, we confirm their results only being able to set an upper limit on the Fe-K\(\alpha\) EW. We then performed the same spectral fitting in the E and W nuclear regions. As shown in Table \ref{tab:fits}, due to the lower statistics with respect to the 4.5'' region, neutral Fe-K\(\alpha\) is detected, although with lower significance with respect to Model a, with line fluxes of \({1.78}_{-1.27}^{+1.51}\) and \({2.39}_{-1.27}^{+1.59} \times{10}^{-7}\mbox{ cm}^{-2}\mbox{ s}^{-1}\) for {E} and {W} nucleus, respectively.
\item[Model c)] Then, we added a second line with a rest-frame energy fixed at 6.7 keV (model c) to try to evaluate the relative contribution of Fe-K\(\alpha\) and Fe \textsc{xxv} lines. With this model the neutral iron line is detected only in the W nucleus region with a flux accounting \(\sim 25\%\pm 10\%\) of the total line flux, while both in the E nuclear and in the central 4.5" region we can only put an upper limit of \(\sim 5\times {10}^{38}\mbox{ erg}\mbox{ s}^{-1}\) on the Fe-K\(\alpha\) emission.

To further test the possible contribution of a neutral iron emission line to the 6-7 keV emission we see in Figure \ref{fig:sources}, we analyzed the archival \textit{XMM-Newton} observations of Arp 220 discussed in \citealt{2005MNRAS.357..565I}. The data were reduced following a standard procedure, analogous to the one described by \citet{2005MNRAS.357..565I}. The results are also in agreement: in a continuum plus single line model we obtain a best fit peak rest-frame energy \(E=6.65\pm 0.04\mbox{ keV}\). However, if we fit the data with two lines with fixed peak rest-frame energies \(E_1=6.4\mbox{ keV}\) and \(E_2=6.7\mbox{ keV}\), we obtain the results shown in Figure \ref{fig:contours}: a neutral component accounting for up to 40\% of the observed line flux cannot be ruled out at a 90\% confidence level, which is compatible with the fluxes obtained with fixed 6.4 keV lines.
\end{itemize}

\subsection{Thermal Model}

Finally, we investigate the possibility for the Fe-K lines to arise from thermal gas emission - possibly from merged supernova (SN) ejecta and stellar winds present during a starburst. To this end we fitted the 2-8 keV spectra with a collisionally ionized plasma component \textsc{APEC} and an intrinsic absorption component \textsc{ZWABS} at the source redshift (Model d), with element abundances both fixed at solar values and free to vary, and the results are reported in Table \ref{tab:fits}. The temperatures of the gas obtained from the fits are \(\sim 5\mbox{ keV}\), with column densities \(\sim 5\times{10}^{22}\mbox{ cm}^{-2}\) for the W and E regions, and \(\sim 2\times{10}^{22}\mbox{ cm}^{-2}\) in the central 4.5'' region. The metallicity are found to be \({4.58}_{-0.72}^{+1.29}\), \({1.14}_{-0.44}^{+0.75}\) and \({1.62}_{-0.52}^{+0.59}\) in the {E, W} and central 4.5'' regions, respectively. We notice, in particular, that the high metallicity obtained in the eastern nucleus can be due to chemical enrichment by Type II SNe in a starburst region, producing substantial \(\alpha\)-elements but a relatively small amount of iron. The Fe metallicity in this region is very large - about four times solar - but this can be the case for a region with intense star formation (e.g., \citealt{2004ApJ...605L..21F}). We note, however that this model is disfavored by the large statistics shown in Table \ref{tab:fits} indicating that, if present, the thermal gas emission is likely to be sub-dominant with respect to harder emissions, possibly of nuclear origin. The results from model d must be therefore interpreted as upper limits on thermal gas emission.

\section{Discussion}\label{discussion}

Making use of deep \textit{Chandra}-ACIS observation and sub-pixel binning in narrow spectral bands we have detected {three} sources of emission in the 6-7 keV band, {with the westernmost and easternmost ones} coincident with the IR (W) and radio (E) nuclei of Arp 220 (see Figure \ref{fig:sources}). The spectral analysis (Section \ref{spectral_analysis}) showed Fe-K lines with large (\(\gtrsim 1 \mbox{ keV}\)) EWs and with the rest-frame line energies larger than 6.4 keV of the neutral Fe-K\(\alpha\) line, and compatible with 6.7 keV Fe \textsc{xxv} emission.

\subsection{CT AGN Models}
 
In this section we try to constrain the presence of a dual CT AGN in the Arp 220 nuclei. CT AGNs are characterized in the X rays by a hard high energy continuum, a ``reflection" flat continuum in the \(\sim 2-10\) keV range, and a high EW (\(\gtrsim 1\)) keV 6.4 Fe-K\(\alpha\) line \citep[e.g.,][]{1997A&A...325L..13M, 2000MNRAS.318..173M}. Examples of this merger-driven evolution are given by the pairs of nuclei discovered in the 6.4 keV Fe-K line with \textit{Chandra} in the merger infrared  (IR) luminous galaxy NGC 6240 \citep{2003ApJ...582L..15K}.

As for NGC 6240 \citep{2003ApJ...582L..15K}, Arp 220 is a highly disturbed system of galaxies engaged in a major merging interaction. The physical projected separation of the CT nuclei is \(\sim 670\) pc in NGC 6240 and \(\sim 380\) pc in Arp 220, suggesting that the latter may be in a more advanced stage of merging.

The spectral analysis of the individual W and E nuclei, results in the detection of Fe emission lines. The rest-frame energy of the line, however, is larger than the 6.4 keV of the K\(\alpha\) line and suggests a contribution from 6.7 keV shock-ionized Fe \textsc{xxv} line. However, as shown in Figure \ref{fig:contours}, if we \textit{assume} that both 6.4 keV and 6.7 keV lines are present in the spectrum, we obtain an acceptable fit to the \textit{XMM-Newton} data which allow for 40\% 6.4 keV contribution. The statistics, however, does not allow us to disentangle the 6.4 and 6.7 keV line contribution to the observed emission in \textit{Chandra} data.

If we then consider that 40\% of the Fe-K line flux we estimate from \textit{Chandra} spectra is due to Fe-K\(\alpha\) neutral 6.4 keV emission line, the 2-10 keV \textit{emitted} luminosities inferred from the Fe-K luminosities are \(1.6\times{10}^{42}\mbox{ erg}\mbox{ s}^{-1}\) (E) and \(0.5\times{10}^{42}\mbox{ erg}\mbox{ s}^{-1}\) (W). We note that these corrections are calibrated on ``standard" obscured Seyfert galaxies, with an X-ray reflection efficiency of a few percent \citep{2006ApJ...648..111L}. Hard X-ray observation of ULIRGs have demonstrated that on average this efficiency is much lower for these sources \citep{2009ApJ...691..261T, 2011MNRAS.415..619N}. Consequently, the intrinsic X-ray luminosity of the two AGN detected here could be significantly higher. Considering the values for a standard reflection efficiency, the inferred X-ray luminosity is at least a factor of 3 higher than that expected from a pure starburst with the bolometric luminosity of Arp 220 \citep{2003A&A...399...39R}.

The two nuclei have radio fluxes of 92.4 (E) and 114.6 (W) mJy at 4.7 GHz \citep{2015ApJ...799...10B}, and X-ray to optical slopes \(\alpha({ox}) \approx 2.75\). From the inferred 2-10 keV emitted luminosities of the two nuclear sources we evaluate bolometric luminosities assuming typical AGN SEDs \citep{1994ApJS...95....1E, 2002ApJ...565L..75E}, and the X-ray reflection efficiency of Seyfert galaxies. The estimated AGN bolometric luminosities, which should be regarded as lower limits, are \(\sim 8.3\times{10}^{43}\mbox{ erg}\mbox{ s}^{-1}\) (E) and \(\sim 2.5\times{10}^{43}\mbox{ erg}\mbox{ s}^{-1}\) (W). These represent only a \(\sim 1\%\) of Arp 220 bolometric luminosity \(\sim 6\times {10}^{45}\mbox{ erg}\mbox{ s}^{-1}\) \citep{1988ApJ...325...74S}, confirming that, overall, the emission of Arp 220 is dominated by the starburst component. We note that the AGN luminosity evaluated with IR data by \citet{2009ApJS..182..628V, 2010MNRAS.405.2505N} is much higher than that inferred from our X-ray analysis, which is suggestive of heavy or nearly total obscuration. Lower limits on associated BH masses can be evaluated assuming Eddington limited accretion (with a standard 10\% accretion rate to luminosity conversion efficiency), yielding \(M \sim 5\times{10}^5 M_{\astrosun}\) (E) and \(\sim 1\times{10}^5 M_{\astrosun}\) (W).

\begin{figure}
\centering
\includegraphics[scale=0.31]{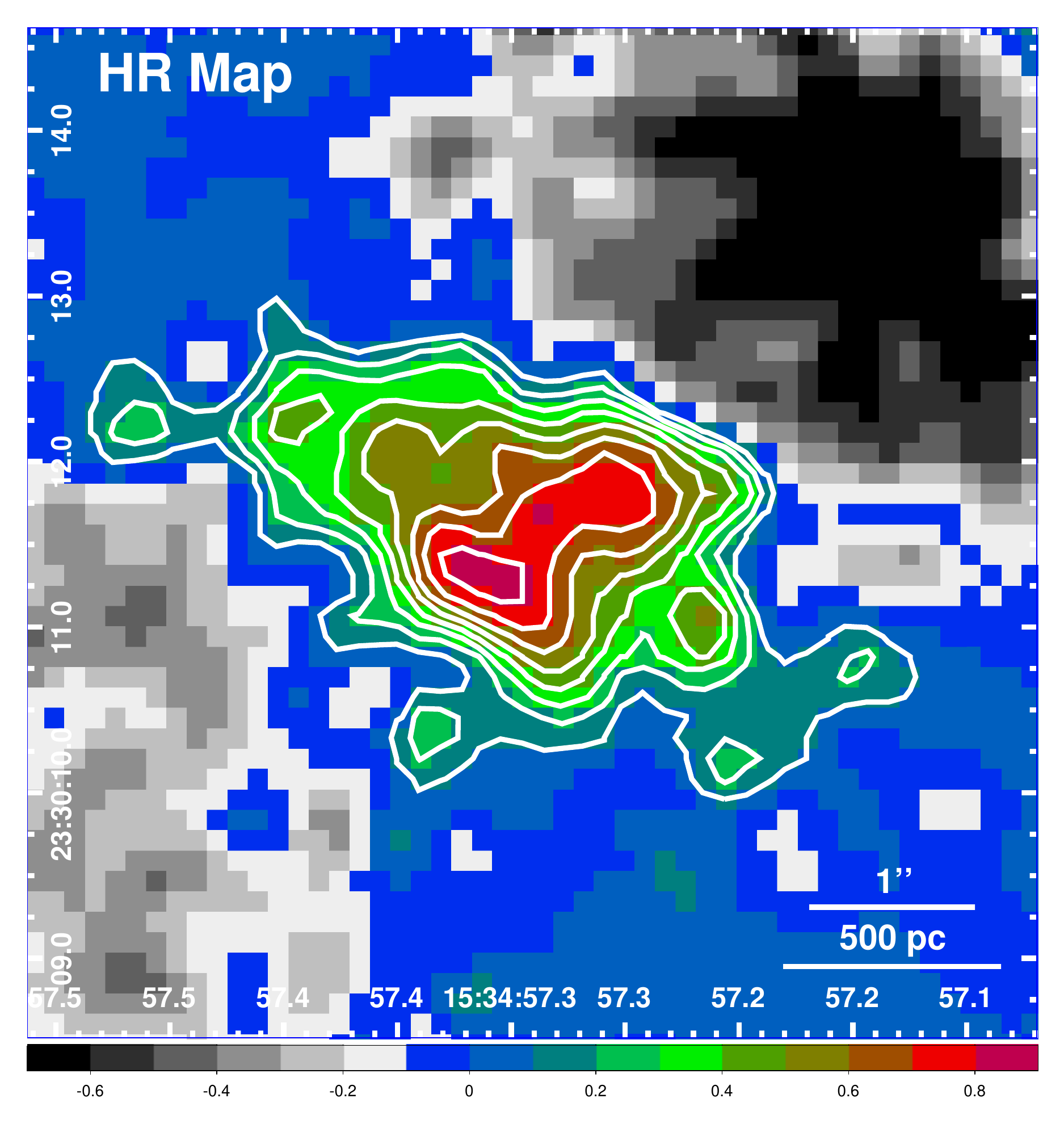}
\includegraphics[scale=0.31]{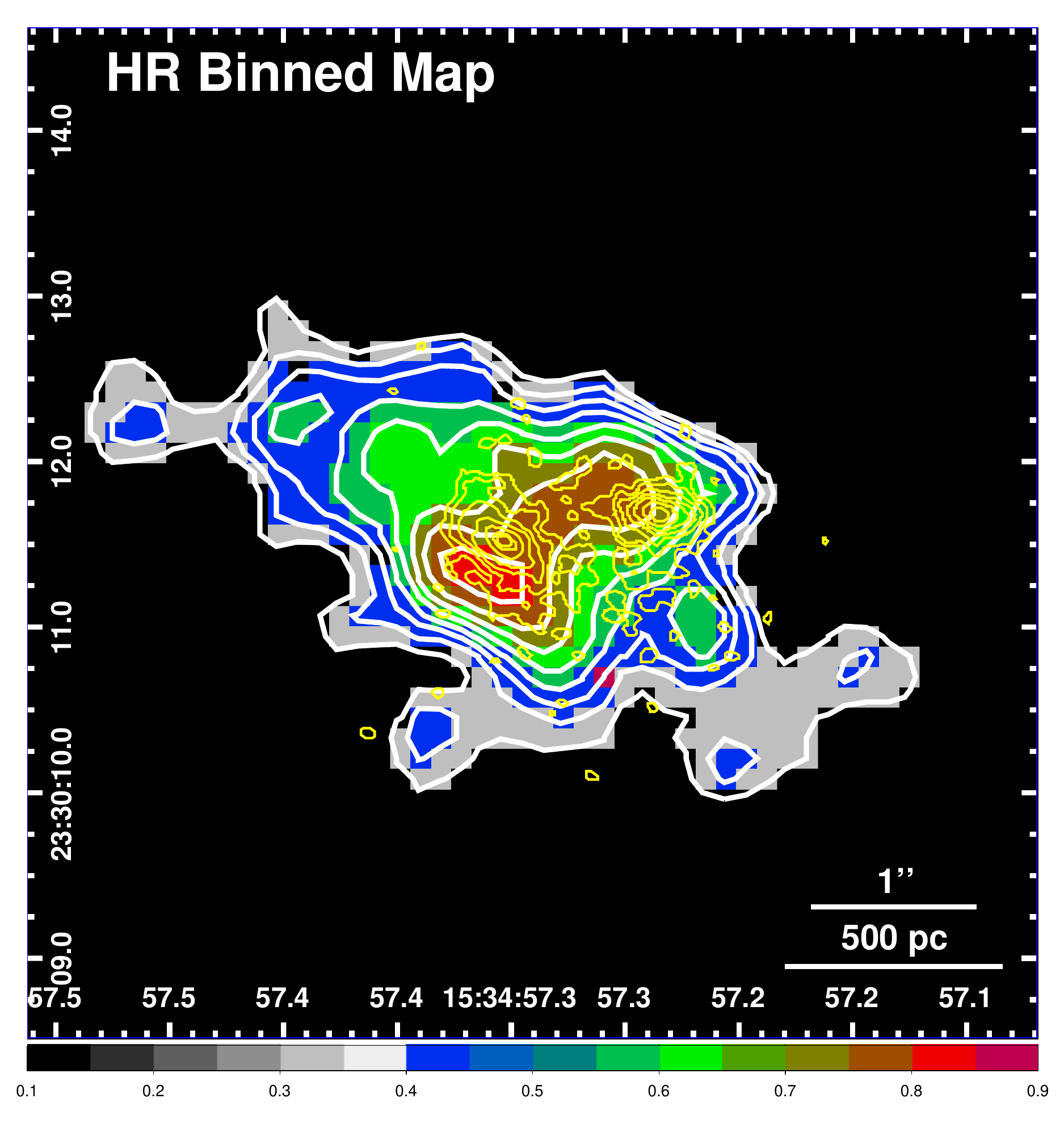}
\includegraphics[scale=0.31]{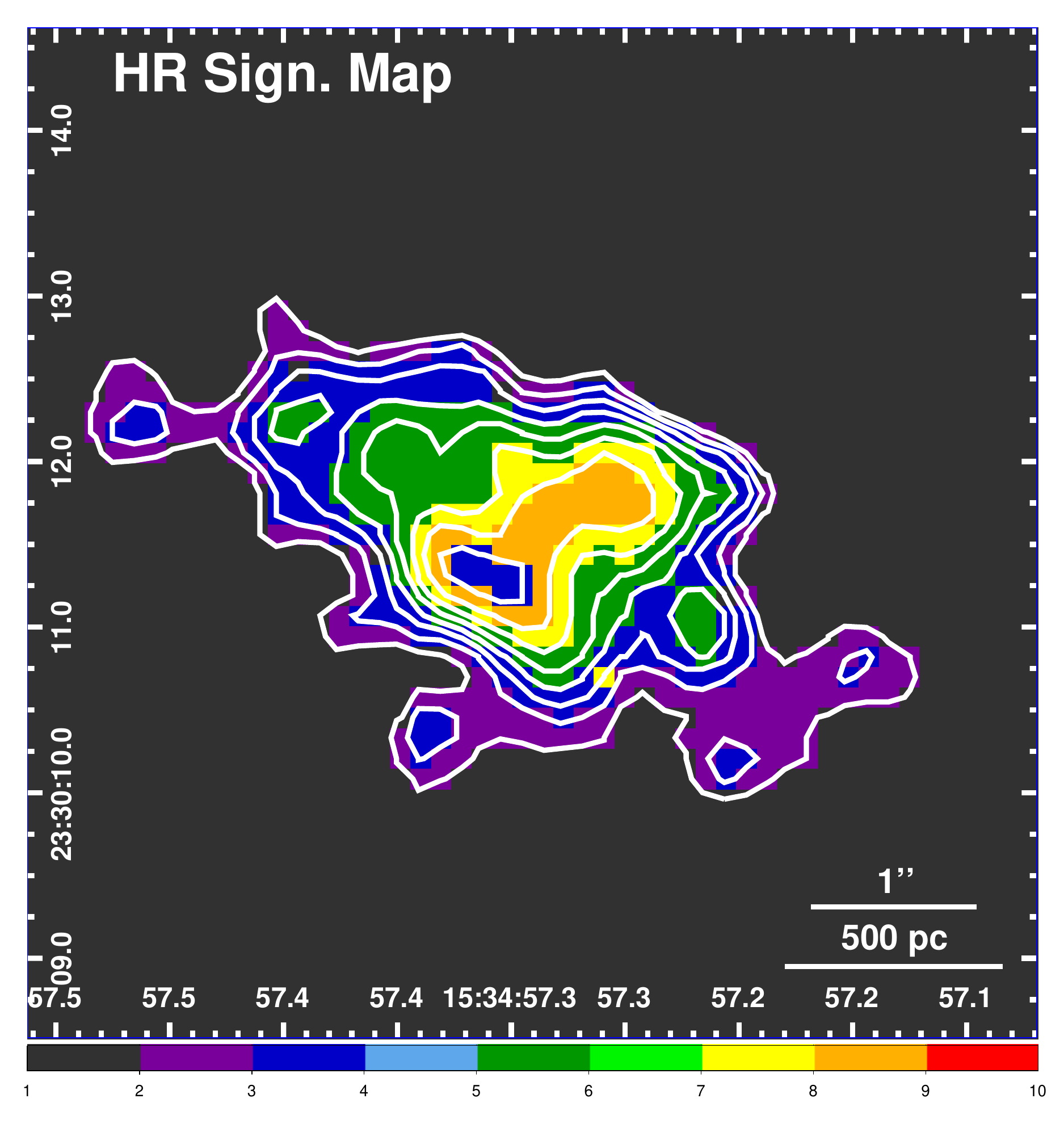}
\caption{(Left Panel) HR map for the central region of Arp 220 with sub-pixel binning 1/4 of the native pixel size and a 3X3 pixel FWHM gaussian filter smoothing. The white contours indicate levels of HR from 0.1 to 0.8 with increments of 0.1. (Central Panel) Same as left panel, but with the map binned using the contour levels. In yellow we show the 33 GHz VLA contours. (Right Panel) Significance map of the HR binned map.}\label{fig:HR_sign}
\end{figure}

\begin{figure}
\centering
\includegraphics[scale=0.47]{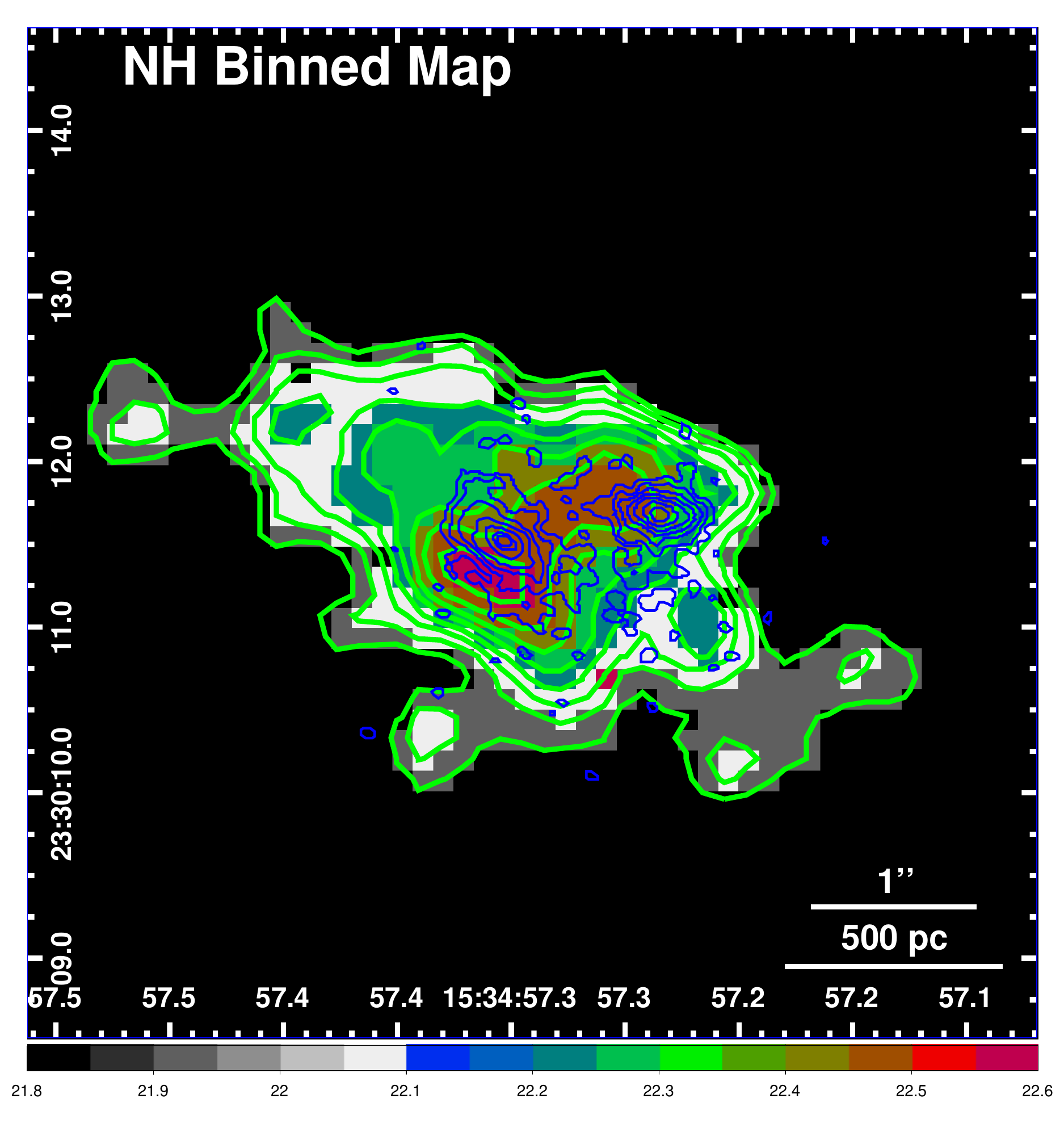}
\includegraphics[scale=0.47]{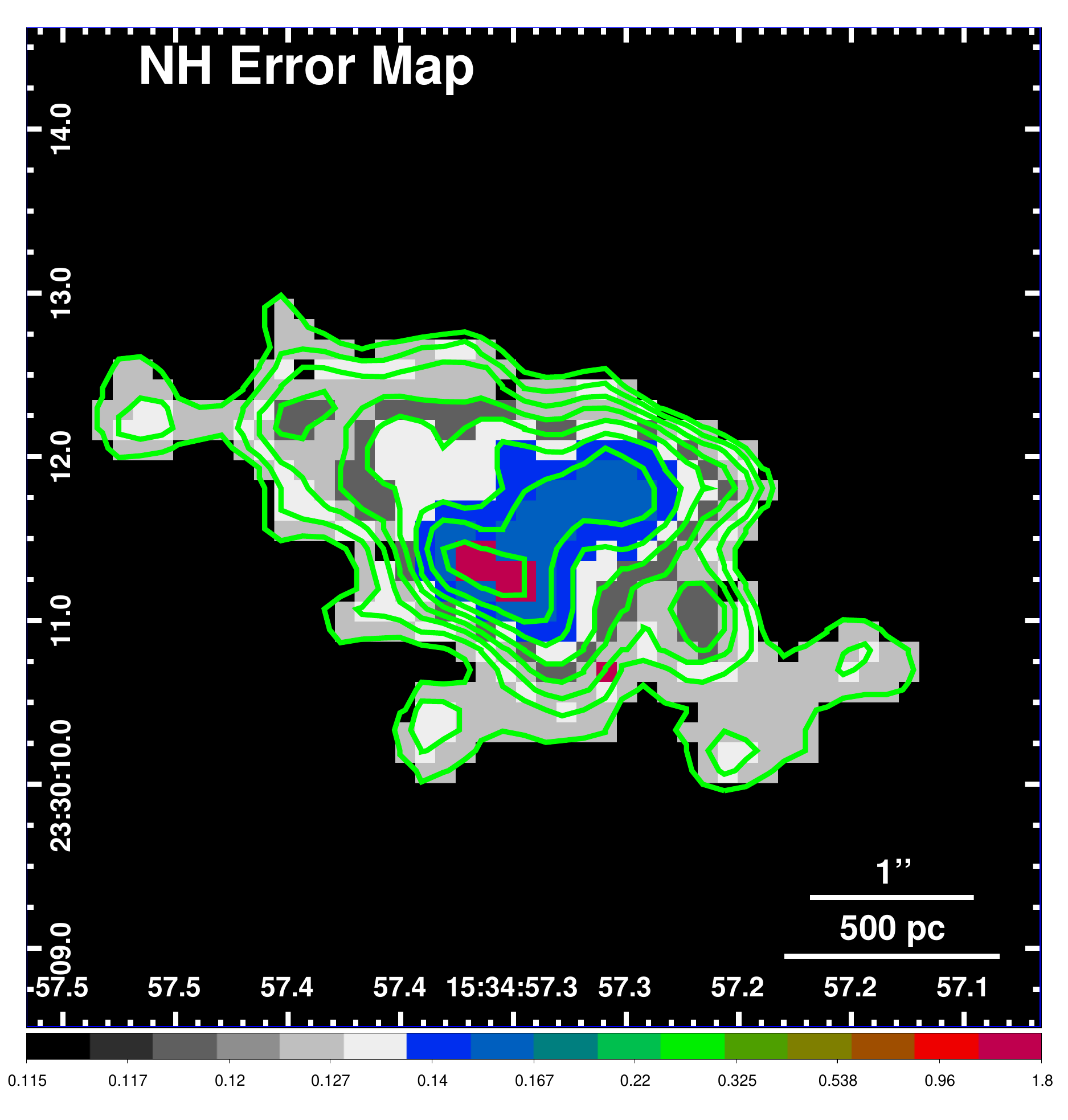}
\caption{(Left Panel) Logarithmic map of the absorption column as evaluated from the HR map presented in the central panel of Figure \ref{fig:HR_sign} assuming a power-law spectrum with 1.8 slope. In blue are overplotted the 33 GHz VLA contours. (Right Panel) Map of the error on \(\log{(N_H/\mbox{cm}^{2})}\) evaluated from the uncertainty on the HR.}\label{fig:NH_sign}
\end{figure}

The lack of 6.4 keV emission may result from a limited visibility of the inner torus surrounding the nuclear AGNs, allowing only the emission of ionized gas to be visible \citep{2005MNRAS.357..565I}. To test this possibility we produced a hardness ratio (HR) map of the central region of this source (Figure \ref{fig:HR_sign}, right panel). This map has been obtained from the event maps in the soft (S) 0.3-2 keV and hard (H) 2-8 keV bands with sub-pixel binning 1/4 of the native pixel size, evaluating \(HR=(H-S)/(H+S)\) and then applying a 3X3 pixel FWHM gaussian filter smoothing. The white contours indicate levels of HR from 0.1 to 0.8 with increments of 0.1. We then binned the event maps using these contours, producing the HR binned map presented in the central panel of Figure \ref{fig:HR_sign}. In this panel we superimpose in yellow the 33 GHz VLA contours from \citet{2015ApJ...799...10B} to highlight the ``bridge" of \(HR\sim 0.8\) that connects the two nuclei detected both in radio and X-ray (see Figure \ref{fig:deconvolution}) and corresponding with X4 source reported by \citet{2002ApJ...581..974C}. On the right panel of Figure \ref{fig:HR_sign} we present a significance map of the HR binned map, evaluated as the ratio between the uncertainty on the HR and the HR itself, showing that this ``bridge" feature is significant at 8 \(\sigma\) level. We then produced simulated spectra, assuming a power-law spectrum with slope 1.8 (as appropriate for AGN emission) and, in addition to the Galactic absorption, an intrinsic absorption component at the source redshift, and evaluated the observed HR such spectra would yield. In this way we converted the binned HR map to the intrinsic absorption column map presented in the left panel of Figure \ref{fig:NH_sign}, where the logarithmic values of \(N_H /\mbox{cm}^{-2}\) are presented, with overplotted in blue the 33 GHz VLA contours. On the right panel of Figure \ref{fig:NH_sign} we show the map of the corresponding error on \(\log{(N_H/\mbox{cm}^{2})}\) evaluated from the uncertainty on the HR. We see that in the region connecting the two nuclei we reach \(N_H \sim {10}^{22.5}\mbox{ cm}^{2}\), similar to that obtained from the spectral fitting with model d, but about four order of magnitudes lower than the value of \(N_{H_2}=2.6\times {10}^{26}\mbox{ cm}^{2}\) reported by \citet{2017ApJ...836...66S} for the W nucleus, {indicating again that the AGN contribution to X-ray emission represents a sub-dominant component with respect to star formation activity (see next section)}.

\subsection{Star Formation Activity}

The nuclear region of Arp 220 is site of an intense star formation with \(\mbox{SFR}\sim 340\,M_{\astrosun}\mbox{ yr}^{-1}\) \citep{2007IAUS..242..437B, 2015ApJ...799...10B}. The X-ray expected luminosity from X-ray binaries, evaluated using the correlation between the galaxy 2-10 keV luminosity and the SFR provided by the \textit{Chandra} survey of LIRGs \citep{2010ApJ...724..559L}, exceeds the observed luminosity by one order of magnitude. It is however possible, as suggested by the previous discussion, that X-ray binaries in Arp 220 are located in compact star-forming regions buried under thick absorption columns with \(N_H >> {10}^{23}\mbox{ cm}^{-2}\) that dims the emission in the 2-10 keV band \citep{2010ApJ...724..559L}.

As discussed before, the Fe \textsc{xxv} emission may be related with the thermal gas from merged SN ejecta and stellar winds present during a starburst, as exemplified by the extended Fe \textsc{xxv} regions of the merger-dual CT AGN NGC 6240 \citealt{2014ApJ...781...55W}). Extended Fe xxv line emission is also observed in other well-known starburst galaxies like NGC 253 and M82 (e.g., \citealt{2001A&A...365L.174P, 2011ApJ...742L..31M}), the merger-dual CT AGN NGC 6240 \citealt{2014ApJ...781...55W}, and the integrated spectrum of a number of LIRG/ULIRG systems \citep{2009ApJ...695L.103I}, suggesting the existence of similar high temperature plasma. 

According to the starburst-driven superwind model, a hot gas bubble of internally shocked wind material with a temperature of several keV forms in the region of intense star formation \citep{1985Natur.317...44C, 1994ApJ...430..511S, 2007ApJ...658..258S}; this hot gas eventually flows outward as a high-speed (few \(1000\mbox{ km}\mbox{ s}^{-1}\)) wind. As a comparison, we note that the observed diffuse hard X-ray emission in Arp 220 has a luminosity one order of magnitude higher than that observed in the classic superwind system M82 (\(L_{2-10\mbox{ keV}} = 4 \times{10}^{39}\mbox{ erg}\mbox{ s}^{-1}\); \citealt{2007ApJ...658..258S}).
Following the calculation of \citet{2014ApJ...781...55W} we can evaluate the shock velocity \(\varv = \sqrt{16\,k T/3\mu}\) (where \(\mu\) is the mean mass per particle and \(k\) is Boltzmann’s constant) from the observed gas temperature \(\sim 5\mbox{ keV}\), obtaining \(\varv \simeq 2000 \mbox{ km}\mbox{ s}^{-1}\). Such velocities are unlikely to be due to shocks induced by merging process of galaxies that collide with velocities of the order of hundreds of \(\mbox{ km}\mbox{ s}^{-1}\) even in the case of direct collision systems \citep{2003A&A...408L..13B}, and they are as well larger than the velocity estimated from the CO line width of \(500-600\mbox{ km}\mbox{ s}^{-1}\) \citep{2008ApJ...684..957S, 2017ApJ...836...66S}.

To further investigate whether the thermal energy content of the hot gas could be powered by thermalization of SNe shocks, we compare the thermal gas energy with the kinetic energy input from the SNe during the starburst. Using the Fe \textsc{xxv} line emission as a tracer for the extent of the hottest thermal component, and following the method of \citet{2010ApJ...723.1375R}, from the emission measure we obtain - assuming a filling factor of 1\%- the hot gas masses of \(2.6\times{10}^5\,M_{\astrosun}\) and \(6.6\times{10}^5\,M_{\astrosun}\) for the W and E nuclei, respectively. For solar abundances this yields an iron mass of \(6.4\times{10}^2\,M_{\astrosun}\) and \(1.7\times{10}^3\,M_{\astrosun}\) for the W and E nuclei, respectively. From the temperatures we get from the spectral fit we obtain a thermal energy of \(E_{th,W}={1.3}\times{10}^{55}\mbox{ erg}\) and \(E_{th,E}={3.1}\times{10}^{55}\mbox{ erg}\), for the W and E nuclei, respectively. Assuming a SN rate for Arp 220 of \(4\mbox{ yr}^{-1}\) \citep{2006ApJ...647..185L} and that 10\% of the kinetic energy input (\({10}^{51}\mbox{ erg}\) per SN) is converted into thermal energy of the hot gas \citep{1985Natur.317...44C, 1998ApJ...500...95T}, we obtain that the total energy deposited during the past starburst is \(\sim 4\times{10}^{50}\mbox{ erg}\mbox{ yr}^{-1}\). \citet{2000ApJ...537..613A} proposed for Arp 220 a star formation model consisting of multiple starbursts of very high SFR (\(\sim {10}^3\,M_{\astrosun}\mbox{ yr}^{-1}\)) and short duration (\(\sim {10}^5\mbox{ yr}\)). With this starburst period we obtain that the gas thermal energy is comparable with the kinetic energy input from the SNe. On the other hand, the cooling time of this thermal gas is \(t_{cool} = E_{th}/L_x \sim 50\) Myr, which is far larger than the starburst duration, allowing the X-ray bright phases to survive once the starburst ceases. Note that such a short starburst can also explain the lack of X-ray point sources, that with a SFR (\(\sim {10}^3\,M_{\astrosun}\mbox{ yr}^{-1}\)), would attain a luminosity \(\sim {10}^{42}\mbox{ erg}\mbox{ s}^{-1}\) \citep{2012MNRAS.419.2095M}. We can also compare the iron yield SNe over the starburst with the iron mass evaluated from the spectral fitting of the thermal gas emission. Using an ejected iron mass per Type II SN of \(8.4 \times {10}^{-2} \,M_{\astrosun}\) \citep{1999ApJS..125..439I}, a SN of \(4\mbox{ yr}^{-1}\) \citep{2006ApJ...647..185L} over a starburst of \({10}^5\mbox{ yr}\) would yield \(\sim 3.4\times{10}^4\,M_{\astrosun}\) of iron, more than enough to account for the iron mass of \(1.7\times{10}^3\,M_{\astrosun}\) and \(6.4\times{10}^2\,M_{\astrosun}\) evaluated before for the {E} and {W} nuclei, respectively.
{On the other hand, a continuous SFR of \(350\, M_{\astrosun}\mbox{ yr}^{-1}\) over a period of \(\sim 1.5\times{10}^7\mbox{ yr}\) \citep{2000ApJ...537..613A} would yield a SNe kinetic energy input exceeding the gas thermal energy by two order of magnitudes and a cooling time about three time the starburst duration, with a yield of \(\sim 3.4\times{10}^6\,M_{\astrosun}\) of iron.}

These estimates demonstrate that the observed diffuse thermal gas traced by the highly ionized iron line emission in Arp 220 is consistent with being heated by SNe shocks in the starburst, although high velocity winds from hidden AGNs may also be present.

\section{Conclusions}\label{conclusions}

We presented an imaging and spectral analysis of the nuclear region of the ULIRG merger Arp 220 which make use of deep \textit{Chandra}-ACIS observations that sum to \(\sim 300\mbox{ ks}\). Narrow-band spectral imaging with sub-pixel resolution of the innermost nuclear region reveals two Fe-K emitting sources, spatially coincident with the infrared and radio emitting nuclear clusters, and separated by 1'' (\(\sim 380\) pc at a distance of \(\sim\) 80 Mpc). We estimate Fe-K equivalent width \(\sim 1\) keV or possibly greater for both sources, and observed 2-10 keV luminosities \(L_X \sim 2\times{10}^{40}\mbox{ erg}\mbox{ s}^{-1}\) (W) and \(\sim 3 \times {10}^{40}\mbox{ erg}\mbox{ s}^{-1}\) (E).

In the narrow 6-7 keV band the emission from these regions is dominated by the 6.7 keV Fe \textsc{xxv} emission line, suggesting a contribution from collisionally ionized gas or starburst regions \citep[see also][]{2005MNRAS.357..565I}. The nuclear regions appear to be filled with a thermal gas at \(\sim 5\) keV whose energetic content can be accounted for by kinetic energy injection in the interstellar medium by Type II SNe. As NGC 6240 \citep{2014ApJ...781...55W}, the thermal gas surrounding the nuclei and responsible for the hard X-ray emission has a thermal energy comparable with the kinetic energy injected in the surrounding medium by Type II SNe in a short (\(\sim {10}^5\mbox{ yr}\)) starburst episode.

{The X-ray emission from the eastern nucleus appears morphologically coincident with the disk emission as mapped by 2.6 mm continuum ALMA and CO 2-1 IRAM PdBi data, suggesting starburst dominance in this region. However, \textit{Chandra} data allow us to constrain the contribution from (possibly dust-obscured) AGNs. In fact,} our analysis of the \textit{XMM-Newton} data confirms the presence of Fe \textsc{xxv} emission line but allows 40\% of the narrow-band emitted flux form the neutral 6.4 keV line. Based on the Fe-K detections, we infer lower limits on the bolometric luminosity of the AGNs in Arp 220 of \(8.3\times{10}^{43}\mbox{ erg}\mbox{ s}^{-1}\) for the E nucleus, and \(\sim 2.5\times{10}^{43}\mbox{ erg}\mbox{ s}^{-1}\) for the W nucleus. These are a few percent of the total ULIRG bolometric luminosity, confirming that overall the emission of this source is dominated by the starburst component, as estimated from the mid-IR spectroscopy \citep{2009ApJS..182..628V, 2010MNRAS.405.2505N}. Our results are consistent with previous multi-wavelength indications of nuclear activity in Arp 220 (see Section \ref{intro}), and strengthen the evolutionary association of merging and nuclear activity in galaxies \citep[e.g.,][]{2008ApJ...682L..13H, 2012ApJ...748L...7V}.

These results have only been possible because of the unmatched \textit{Chandra}-ACIS spatial resolution coupled with the use of sub-pixel imaging in narrow spectral bands, which allow us to perform narrow-band, high resolution imaging that gives us a clear picture of the nuclear surroundings.

\acknowledgments
{We acknowledge useful comments and suggestions by our anonymous referee.}
This work is supported by NASA grant G04-15107X (PI. Fabbiano). {JW acknowledges support from National Key Program for Science and Technology Research and Development 2016YFA0400702, and the NSFC grants 11473021,11522323.} We acknowledge support from the CXC, which is operated by the Smithsonian Astrophysical Observatory (SAO) for and on behalf of NASA under Contract NAS8-03060. This research has made use of data obtained from the \textit{Chandra} Data Archive, and software provided by the CXC in the application packages CIAO and Sherpa. This research has made use of Iris software provided by the US Virtual Astronomical Observatory, which is sponsored by the National Science Foundation and the National Aeronautics and Space Administration.

\newpage

\begin{table}
\begin{center}
\caption{Summary of \textit{Chandra} observations of Arp 220.}\label{tab:observations}
\begin{tabular}{|c|c|c|c|c|}
\hline
\hline
 OBSID & DATE & EXPOSURE & PI & 0.5-8 keV COUNTS \\
       &      & (ks)     &    & (central 5'' region) \\ 
\hline
 00869 & 2000-Jun-24 & 57  & Clements & 583  \\
 16092 & 2014-Arp-30 & 170 & Fabbiano & 1390 \\
 16093 & 2014-Jun-24 & 67  & Fabbiano & 543  \\
\hline
\hline
\end{tabular}
\end{center}
\end{table}

\begin{table}
\begin{center}
\caption{Best fit parameters for the extraction regions considered in the main text.}\label{tab:fits}
\begin{tabular}{|l|cccccc|cccccc|cccccc|}
\hline
\hline
& \multicolumn{6}{c|}{W region} & \multicolumn{6}{c|}{E region} & \multicolumn{6}{c|}{Central 4.5''}\\
\hline
Band & \multicolumn{18}{c|}{Net counts (error)} \\
\hline
3-8 keV & \multicolumn{6}{c|}{75(9)} & \multicolumn{6}{c|}{221(15)} & \multicolumn{6}{c|}{659(26)} \\
6-7 keV & \multicolumn{6}{c|}{15(3)} & \multicolumn{6}{c|}{41(6)} & \multicolumn{6}{c|}{90(10)}\\
6-6.4 keV & \multicolumn{6}{c|}{6(2)} & \multicolumn{6}{c|}{11(3)} & \multicolumn{6}{c|}{25(5)}\\
6.4-6.7 keV & \multicolumn{6}{c|}{8(3)} & \multicolumn{6}{c|}{27(5)} & \multicolumn{6}{c|}{55(7)}\\
\hline
Model parameter & \multicolumn{18}{c|}{Best-fit value} \\
\hline
Spectral model (Power-law + lines) & \multicolumn{2}{c}{(a)} & \multicolumn{2}{c}{(b)} & \multicolumn{2}{c|}{(c)} & \multicolumn{2}{c}{(a)} & \multicolumn{2}{c}{(b)} & \multicolumn{2}{c|}{(c)} & \multicolumn{2}{c}{(a)} & \multicolumn{2}{c}{(b)} & \multicolumn{2}{c|}{(c)} \\
\hline
\(\Gamma_{PL}\) & \multicolumn{2}{c}{\({3.40}^*\)} & \multicolumn{2}{c}{\({3.40}^*\)} & \multicolumn{2}{c|}{\({3.40}^*\)} & \multicolumn{2}{c}{\({2.39}^*\)} & \multicolumn{2}{c}{\({2.39}^*\)} & \multicolumn{2}{c|}{\({2.39}^*\)} & \multicolumn{2}{c}{\({3.12}^*\)} & \multicolumn{2}{c}{\({3.12}^*\)} & \multicolumn{2}{c|}{\({3.12}^*\)} \\
\(F_{PL}\, ({10}^{-4}\mbox{ cm}^{-2}\mbox{ s}^{-1}\mbox{ keV}^{-1})\) & \multicolumn{2}{c}{\({18.30}_{-1.72}^{+2.27}\)} & \multicolumn{2}{c}{\({18.97}_{-1.93}^{+2.07}\)} & \multicolumn{2}{c|}{\({18.06}_{-1.90}^{+2.04}\)} & \multicolumn{2}{c}{\({4.48}_{-0.67}^{+0.40}\)}  & \multicolumn{2}{c}{\({5.56}_{-0.51}^{+0.55}\)}  & \multicolumn{2}{c|}{\({4.47}_{-0.50}^{+0.38}\)}  & \multicolumn{2}{c}{\({38.65}_{-3.42}^{+2.43}\)} & \multicolumn{2}{c}{\({44.42}_{-2.26}^{+2.37}\)} & \multicolumn{2}{c|}{\({38.91}_{-2.27}^{+2.29}\)} \\
\(E_{Fe-K}\) (keV, rest-frame) & \multicolumn{2}{c}{\({6.61}_{-0.07}^{+0.06}\)} & \multicolumn{2}{c}{\(6.4^*\)} &  \multicolumn{2}{c|}{\(6.4^*\)} & \multicolumn{2}{c}{\({6.67}_{-0.03}^{+0.03}\)} & \multicolumn{2}{c}{\(6.4^*\)} & \multicolumn{2}{c|}{\(6.4^*\)} & \multicolumn{2}{c}{\({6.65}_{-0.02}^{+0.02}\)} & \multicolumn{2}{c}{\(6.4^*\)} & \multicolumn{2}{c|}{\(6.4^*\)} \\
\(F_{Fe-K}\,({10}^{-7}\mbox{ cm}^{-2}\mbox{ s}^{-1})\) & \multicolumn{2}{c}{\({4.01}_{-1.59}^{+1.97}\)} & \multicolumn{2}{c}{\({2.39}_{-1.27}^{+1.59}\)} & \multicolumn{2}{c|}{\({1.16}_{-1.09}^{+1.42}\)} & \multicolumn{2}{c}{\({9.43}_{-1.91}^{+2.17}\)} & \multicolumn{2}{c}{\({1.78}_{-1.27}^{+1.51}\)} & \multicolumn{2}{c|}{\(<0.86\)} & \multicolumn{2}{c}{\({11.92}_{-2.07}^{+2.21}\)} & \multicolumn{2}{c}{\(<5.01\)} & \multicolumn{2}{c|}{\(<0.88\)} \\
\(L_{Fe-K}\,({10}^{40}\mbox{ erg}\mbox{ s}^{-1})\) & \multicolumn{2}{c}{\({0.32}_{-0.14}^{+0.13}\)} & \multicolumn{2}{c}{\({0.19}_{-0.10}^{+0.11}\)} & \multicolumn{2}{c|}{\({0.09}_{-0.07}^{+0.07}\)} & \multicolumn{2}{c}{\({0.76}_{-0.15}^{+0.17}\)} & \multicolumn{2}{c}{\({0.14}_{-0.08}^{+0.11}\)} & \multicolumn{2}{c|}{\(<0.04\)} & \multicolumn{2}{c}{\({0.96}_{-0.17}^{+0.17}\)} & \multicolumn{2}{c}{\(<0.29\)} & \multicolumn{2}{c|}{\(<0.06\)} \\
\({EW}_{Fe-K}\) (keV) & \multicolumn{2}{c}{\({1.22}_{-0.53}^{+0.57}\)} & \multicolumn{2}{c}{\({0.67}_{-0.36}^{+0.43}\)} & \multicolumn{2}{c|}{\({0.35}_{-0.28}^{+0.29}\)} & \multicolumn{2}{c}{\({1.76}_{-0.43}^{+0.45}\)} & \multicolumn{2}{c}{\({0.27}_{-0.15}^{+0.23}\)} & \multicolumn{2}{c|}{\(<0.11\)} & \multicolumn{2}{c}{\({1.02}_{-0.19}^{+0.19}\)} & \multicolumn{2}{c}{\(<0.27\)} & \multicolumn{2}{c|}{\(<0.06\)} \\
\(E_{Fe-K}\) (keV, rest-frame) & \multicolumn{2}{c}{-} & \multicolumn{2}{c}{-} & \multicolumn{2}{c|}{\(6.7^*\)} & \multicolumn{2}{c}{-} & \multicolumn{2}{c}{-} & \multicolumn{2}{c|}{\(6.7^*\)} & \multicolumn{2}{c}{-} & \multicolumn{2}{c}{-} & \multicolumn{2}{c|}{\(6.7^*\)} \\
\(F_{Fe-K}\,({10}^{-7}\mbox{ cm}^{-2}\mbox{ s}^{-1})\) & \multicolumn{2}{c}{-} & \multicolumn{2}{c}{-} & \multicolumn{2}{c|}{\({3.33}_{-1.58}^{+1.94}\)} & \multicolumn{2}{c}{-} & \multicolumn{2}{c}{-} & \multicolumn{2}{c|}{\({9.62}_{-1.99}^{+2.16}\)} & \multicolumn{2}{c}{-} & \multicolumn{2}{c}{-} & \multicolumn{2}{c|}{\({11.86}_{-2.03}^{+2.22}\)} \\
\(L_{Fe-K}\,({10}^{40}\mbox{ erg}\mbox{ s}^{-1})\) & \multicolumn{2}{c}{-} & \multicolumn{2}{c}{-} & \multicolumn{2}{c|}{\({0.27}_{-0.13}^{+0.15}\)} & \multicolumn{2}{c}{-} & \multicolumn{2}{c}{-} & \multicolumn{2}{c|}{\({0.78}_{-0.19}^{+0.18}\)} & \multicolumn{2}{c}{-} & \multicolumn{2}{c}{-} & \multicolumn{2}{c|}{\({0.96}_{-0.08}^{+0.12}\)} \\
\({EW}_{Fe-K}\) (keV) & \multicolumn{2}{c}{-} & \multicolumn{2}{c}{-} & \multicolumn{2}{c|}{\({1.04}_{-0.47}^{+0.60}\)} & \multicolumn{2}{c}{-} & \multicolumn{2}{c}{-} & \multicolumn{2}{c|}{\({1.80}_{-0.46}^{+0.48}\)} & \multicolumn{2}{c}{-} & \multicolumn{2}{c}{-} & \multicolumn{2}{c|}{\({0.97}_{-0.08}^{+0.13}\)} \\
C-stat (d.o.f.) & \multicolumn{2}{c}{34.9(29)} & \multicolumn{2}{c}{47.0(30)} & \multicolumn{2}{c|}{42.3(29)} & \multicolumn{2}{c}{86.3(66)} & \multicolumn{2}{c}{137.5(67)} & \multicolumn{2}{c|}{85.5(66)} & \multicolumn{2}{c}{288.1(211)} & \multicolumn{2}{c}{352.2(212)} & \multicolumn{2}{c|}{292.3(211)} \\
\hline
\(L_{2-10\mbox{ keV}}\,({10}^{40}\mbox{ erg}\mbox{ s}^{-1})\) & \multicolumn{2}{c}{\({1.99}_{-0.23}^{+0.20}\)} & \multicolumn{2}{c}{\({1.91}_{-0.23}^{+0.23}\)} & \multicolumn{2}{c|}{\({2.01}_{-0.23}^{+0.26}\)} & \multicolumn{2}{c}{\({3.22}_{-0.32}^{+0.31}\)} & \multicolumn{2}{c}{\({3.08}_{-0.29}^{+0.31}\)} & \multicolumn{2}{c|}{\({3.23}_{-0.33}^{+0.32}\)} & \multicolumn{2}{c}{\({6.81}_{-0.39}^{+0.39}\)} & \multicolumn{2}{c}{\({6.56}_{-0.43}^{+0.44}\)} & \multicolumn{2}{c|}{\({7.13}_{-0.30}^{+0.35}\)} \\
\hline
Model parameter & \multicolumn{18}{c|}{Best-fit value} \\
\hline
Spectral model (\textsc{APEC}) & \multicolumn{6}{c|}{(d)} & \multicolumn{6}{c|}{(d)} & \multicolumn{6}{c|}{(d)} \\
\hline
\(N_H\) (\({10}^{22}\mbox{ cm}^{-2}\)) & \multicolumn{3}{c}{\({5.63}_{-1.07}^{+3.20}\)} & \multicolumn{3}{c|}{\({5.65}_{-1.32}^{+1.89}\)} & \multicolumn{3}{c}{\({5.83}_{-0.69}^{+0.90}\)} & \multicolumn{3}{c|}{\({4.75}_{-0.74}^{+0.74}\)} & \multicolumn{3}{c}{\({2.42}_{-0.51}^{+0.79}\)} & \multicolumn{3}{c|}{\({2.28}_{-0.61}^{+1.08}\)} \\
\(kT\) (keV) & \multicolumn{3}{c}{\({5.58}_{-2.27}^{+5.34}\)} &
\multicolumn{3}{c|}{\({5.44}_{-1.51}^{+2.71}\)} & \multicolumn{3}{c}{\({5.18}_{-1.05}^{+0.79}\)} & \multicolumn{3}{c|}{\({5.53}_{-0.70}^{+1.11}\)} & \multicolumn{3}{c}{\({5.30}_{-1.14}^{+1.26}\)} & \multicolumn{3}{c|}{\({5.21}_{-1.02}^{+1.18}\)} \\
EM \(({10}^{-4}\mbox{ cm}^{-2}\mbox{ s}^{-1})\) & \multicolumn{3}{c}{\({2.95}_{-0.76}^{+1.49}\)} & \multicolumn{3}{c|}{\({2.89}_{-0.72}^{+1.88}\)} & \multicolumn{3}{c}{\({4.58}_{-0.72}^{+1.29}\)} & \multicolumn{3}{c|}{\({2.31}_{-0.42}^{+0.65}\)} & \multicolumn{3}{c}{\({7.42}_{-0.94}^{+1.22}\)} & \multicolumn{3}{c|}{\({6.45}_{-1.15}^{+1.85}\)} \\
Z & \multicolumn{3}{c}{\(1^*\)} & \multicolumn{3}{c|}{\({1.14}_{-0.44}^{+0.75}\)} & \multicolumn{3}{c}{\(1^*\)} & \multicolumn{3}{c|}{\({4.14}_{-1.29}^{+2.30}\)} & \multicolumn{3}{c}{\(1^*\)} & \multicolumn{3}{c|}{\({1.62}_{-0.52}^{+0.59}\)} \\
C-stat (d.o.f.) & \multicolumn{3}{c}{111.6(40)} & \multicolumn{3}{c|}{111.5(39)} & \multicolumn{3}{c}{261.6(94)} & \multicolumn{3}{c|}{247.9(93)} & \multicolumn{3}{c}{306.8(310)} & \multicolumn{3}{c|}{305.7(309)} \\
\hline
\(L_{0.5-8\mbox{ keV}}\,({10}^{40}\mbox{ erg}\mbox{ s}^{-1})\) & \multicolumn{3}{c}{\({3.85}_{-0.62}^{+0.70}\)} & \multicolumn{3}{c|}{\({3.89}_{-1.41}^{+1.51}\)} & \multicolumn{3}{c}{\({5.89}_{-0.89}^{+0.86}\)} & \multicolumn{3}{c|}{\({5.42}_{-1.62}^{+1.52}\)} & \multicolumn{3}{c}{\({9.59}_{-4.46}^{+4.96}\)} & \multicolumn{3}{c|}{\({9.65}_{-3.15}^{+3.55}\)} \\
\hline
\hline
\end{tabular}
\end{center}
\end{table}

\end{document}